\documentclass[fleqn,usenatbib]{mnras}

\usepackage{newtxtext,newtxmath}

\usepackage[T1]{fontenc}

\usepackage{graphicx}	
\usepackage{amsmath}	

\usepackage[english]{babel}
\newcommand{\myr}{$\rm mas\,yr^{-1}$}
\newcommand{\kms}{$\rm km\,s^{-1}$}
\newcommand{\kmskpc}{$\rm km\,s^{-1}\,kpc^{-1}$}
\newcommand{\kmskpcc}{$\rm km\,s^{-1}\,kpc^{-2}$}

\title[A new kinematic model of the Galaxy: analysis of the stellar velocity field] {A new kinematic model of the Galaxy: analysis of the stellar velocity field from \emph{Gaia} Data Release 3}

\author[Akhmetov et al.]{
V. S. Akhmetov$^{1,3}$\thanks{Contact e-mail: \href{mailto:akhmetovvs@gmail.com}{akhmetovvs@gmail.com}(VSA)},
B. Bucciarelli$^{1}$,
M. Crosta$^{1}$,
M. G. Lattanzi$^{1}$,\newauthor
A. Spagna$^{1}$, 
P. Re Fiorentin$^{1}$ and E. Yu. Bannikova$^{2,3,4}$
\\
$^{1}$INAF-Osservatorio Astrofisico di Torino, Via Osservatorio 20, Pino Torinese, Turin, I-10025, Italy\\
$^{2}$INAF -- Astronomical Observatory of Capodimonte, Salita Moiariello 16, I-80131, Naples, Italy\\
$^{3}$V.N.Karazin Kharkiv National University, Svobody sq. 4, 61022, Kharkiv, Ukraine\\
$^{4}$Institute of Radio Astronomy, National Academy of Sciences of Ukraine, Mystetstv 4, Kharkiv, UA-61002, Ukraine
}

\date{Accepted XXX. Received YYY; in original form ZZZ}

\pubyear{2022}

\begin{document}
\label{firstpage}
\pagerange{\pageref{firstpage}--\pageref{lastpage}}
\maketitle

\begin{abstract}
This work presents the results of a kinematic analysis of the Galaxy that 
uses a new model as applied to the newest available  \emph{Gaia} data. We carry out the Taylor decomposition of the velocity field up to second order for 18 million high luminosity stars (i.e. OBAF-type stars, giants and subgiants) from the \emph{Gaia} DR3 data. We determine the components of mean stellar velocities and their first and second partial derivatives (relative to cylindrical coordinates) for more than 28 thousand points in the plane of our Galaxy. We estimate Oort's constants $A$, $B$, $C$, and $K$ and other kinematics parameters and map them as a function of Galactocentric coordinates. The values found confirm the results of our previous works and are in excellent agreement with those obtained by other authors in the Solar neighbourhood. In addition, the introduction of second order partial derivatives of the stellar velocity field allows us to determine the values of the vertical gradient of the Galaxy azimuthal, radial and vertical velocities. Also, we determine the mean of the Galaxy rotation curve for Galactocentric distances from 4 kpc to 18 kpc by averaging Galactic azimuths in the range  \mbox{-30$^\circ$ < $\theta$  < +30$^\circ$} about the direction Galactic Centre -- Sun --Galactic anticentre. Maps of the velocity components and of their partial derivatives with respect to coordinates within 10 kpc of the Sun reveal complex substructures, which provide clear evidence of non-axisymmetric features of the Galaxy. Finally, we show evidence of  differences in the Northern and Southern hemispheres stellar velocity fields. 

\end{abstract}


\section{Introduction}
\label{sec:intro}

The \emph{Gaia}~DR3 catalogue (\cite{Prusti2016, Brown2020, Vallenari2022}) has provided new data for studying the kinematics of stars in the Milky Way. The presence of the spatial positions and velocities of 33 million stars in this release makes it possible to obtain new information about stellar kinematics in our Galaxy. Thus, the large amount of data and the new level of its precision require us to develop new approaches to its analysis in the context of kinematic studies of the Galaxy.

Recently, some new methods are often used to analyse stellar kinematics \citep{Poggio2018, Eilers2019, Antoja2018, Drimmel2022, Nelson2022}. 
\cite{Nelson2022} constructed a non-parametric, smooth and differentiable model of the underlying velocity field using a sparse Gaussian process algorithm based on induction points. They estimated the Oort constants $A$, $B$, $C$ and $K$ and presented the maps of the velocity field and the divergence within 2 kpc of the Sun. \cite{Drimmel2022} selected various stellar populations to explore and identify non-axisymmetric features in the Milky disc and mapped the spiral structures associated with star formation in the scale range 4-5 kpc from the Sun. They used the sample of the red giant branch to obtain the velocity field maps of the Milky Way as far as 8 kpc from the Sun. \cite{Poggio2018} showed that the large-scale kinematics has a clear signature of the warp of the Milky Way appearing as a gradient of 5-6 \kms in the vertical velocities between 8 and 14 kpc of the Galactic radius. 

Using the Ogorodnikov--Milne (O--M) model \cite{Fedorov2021} obtained the kinematic parameters of red giants and subgiants from the \emph{Gaia}~EDR3 catalogue along the direction of the Galactic center -- the Sun -- the Galactic anticentre.
\cite{Fedorov2023} mapped the kinematic parameters in the plane of our Galaxy using the following approach. They solved the equations for the scalar stellar velocity field to find the velocity components of the centroid and their first order partial derivatives with respect to the Galactocentric cylindrical coordinate system. \cite{Dmytrenko2023} analysed  the O--M deformation velocity tensor to determine the vertices of selected stellar samples.

In addition to classical physical models (e.g., the Oort–Lindblad, Ogorodnikov–Milne), other mathematical models are often used to analyze the stellar kinematics of the Galaxy. Assuming that stellar proper motions and radial velocities are the components of the velocity vector field, it is reasonable to use the methods of decomposing the corresponding stellar velocity field into a set of vector spherical harmonics (VSH). This approach allows the detection of all systematic components contained in the studied velocity field.
The comparison of the decomposition coefficients with the model parameters shows whether the models are complete, and allows to reveal any significant systematics that are not included in the models. This approach is very effective, as previous studies have shown \citep{Vityazev2005, Makarov2007,  Mignard2012, Velichko2020}. Nevertheless, the main problem of this approach is to find the physical sense of the decomposition coefficients and to relate them with the kinematics features of our Galaxy.

In this paper, we obtain a detailed kinematics in the Galactic plane based on the \emph{Gaia} DR3 data, which are provided with spatial positions and velocities. We derive the kinematic parameters by analysing with a new method the high luminosity stars, whose velocities are considered relative to the Galactic center. We exclude the velocity of the Sun $V_{\odot}$ relative to the Galactic center from the relative velocities of all stars. Consequently, this allowed us to analyze the field of stellar velocities in a Galactocentric cylindrical coordinate system.

\section{Data}
\label{sec:preparation}

 We select \emph{Gaia} DR3 sources for which the six dimensional phase-space coordinates can be computed, i.e. all stars with five-parameters astrometric solution (sky positions, parallax and proper motions) and line-of-sight (radial) velocity (about 33 million stars in Gaia DR3). We retain only objects that have i) Renormalized Unit Weight Error \texttt{ruwe} $ < 1.4$ in order to discard the sources with problematic astrometric solutions, astrometric binaries, and other anomalous cases \citep{Lindegren2018}, ii) the relative parallax error $\varpi/\sigma_\varpi  > 5$ (i.e. the inverse-parallax distances better than 20\% ), iii) and three \emph{Gaia} photometric bands $G$, $G_{BP}$ and $G_{RP}$. We estimate the distance to the stars as $1/\varpi$, where $\varpi$ is the parallax. As it was noticed in \citet{Antoja2018}, the differences between distances, determined with help of a Bayesian method and of the inverse of the parallax, are between $-2\%$ and $0.6\%$ for 90$\%$ stars with $\varpi/\sigma_\varpi > 5$. Such a result is expected for small relative errors in parallax. 

We also exclude stars that are at large distances from the mid-plane of the Galaxy $|z|>1 $ kpc and also unbound stars or high-velocity stars with Galactocentric velocity $ V_{GC} = \sqrt{V_R^2+V_{\theta}^2+V_Z^2} > 500 $ \kms  \citep{Marchetti2022}.

Furthermore, for our kinematic analysis we select only high luminosity stars with $M_G<4$ and exclude faint stars that are dominant near the Sun and are incomplete at large distances. The spatial distribution of our stellar tracers is more uniform and covers a wide range of Galactocentric distance range, i.e., 0 kpc $< R < $ 18 kpc.

\begin{figure}
   \centering
\resizebox{\hsize}{!}
   {\includegraphics{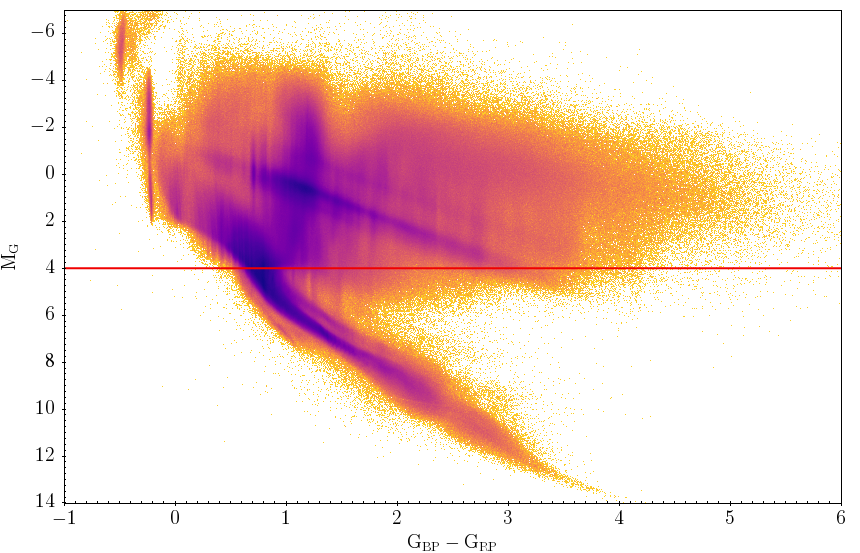}}
  \caption{Selection of high luminosity stars, such as  OBAF-type stars, subgiants and giants on the colour diagram: $G_{BP}-G_{RP}$–absolute magnitude $M_G$. We use stars above the red horizontal line with a value of $M_G$ = 4.}
\label{fig:MS_RG}
\end{figure}

Fig. \ref{fig:MS_RG} shows the CMD (Color-Magnitude Diagrams)  $G_{BP}-G_{RP}$ vs.\ $M_G$ as the density map of all the 27.3 million stars selected with the astrometric, photometric, and spectroscopic quality cuts described in the previous paragraphs. Here, we apply the $A_G$ extinction (\texttt{ag\_gspphot})  and the reddening  $E(G_{BP}-G_{RP})$ (\texttt{ebvminrp\_gspphot}) estimated by the astrophysics solution of the \emph{Gaia} DR3 $Apsis$ from \citet{Creevey2022}. 
This CMD includes also 6.1 million stars (22\%) for which $Apsis$ does not provide extinction and reddening information (i.e.\ \texttt{ag\_gspphot} =0). As in \citep{RuizDern2018}, for these stars we estimate a conservative value of the absolute magnitude by adopting the 99th percentile of the parallax probability density function: $\varpi+2.32\sigma_{\varpi}$.
Such parallax criterion allows us to cover the full red giant branch and avoid the contamination of dwarfs.

The horizontal red line $M_G=4$ in Fig. \ref{fig:MS_RG} splits the objects into the main-sequence stars (below the line) and the high luminosity stars, such as upper main sequence stars, subgiants, and giants.

In summary, for our kinematic analysis, we use 18 million stars that satisfy the following criteria:
\begin{align}
\label{Mabs}
 & \texttt{ruwe} < 1.4 ;\nonumber\\
 & \texttt{non\_single\_star} = 0;\nonumber\\
 & \frac{\varpi}{\sigma_{\varpi}} > 5; \\
 &  |z| < 1 \mathrm{\, kpc}; \nonumber\\
 &  V_{GC} < 500 \mathrm{\, kms ;} \nonumber\\
 &  \begin{cases}
 &  m_G + 5 + 5 \log_{10} \Big(\frac{\varpi}{1000}\Big) - \texttt{ag\_gspphot}< 4 ; \nonumber\\
 &  m_G + 5 + 5 \log_{10} \Big(\frac{\varpi+2.32\sigma_{\varpi}}{1000}\Big)< 4, (\texttt{ag\_gspphot}=0).\nonumber\\
\end{cases}
\end{align}


\section{Coordinate systems and mean velocity distribution}
\label{sec:coordinate_systems}

\begin{figure}
   \centering
\resizebox{\hsize}{!}
   {\includegraphics{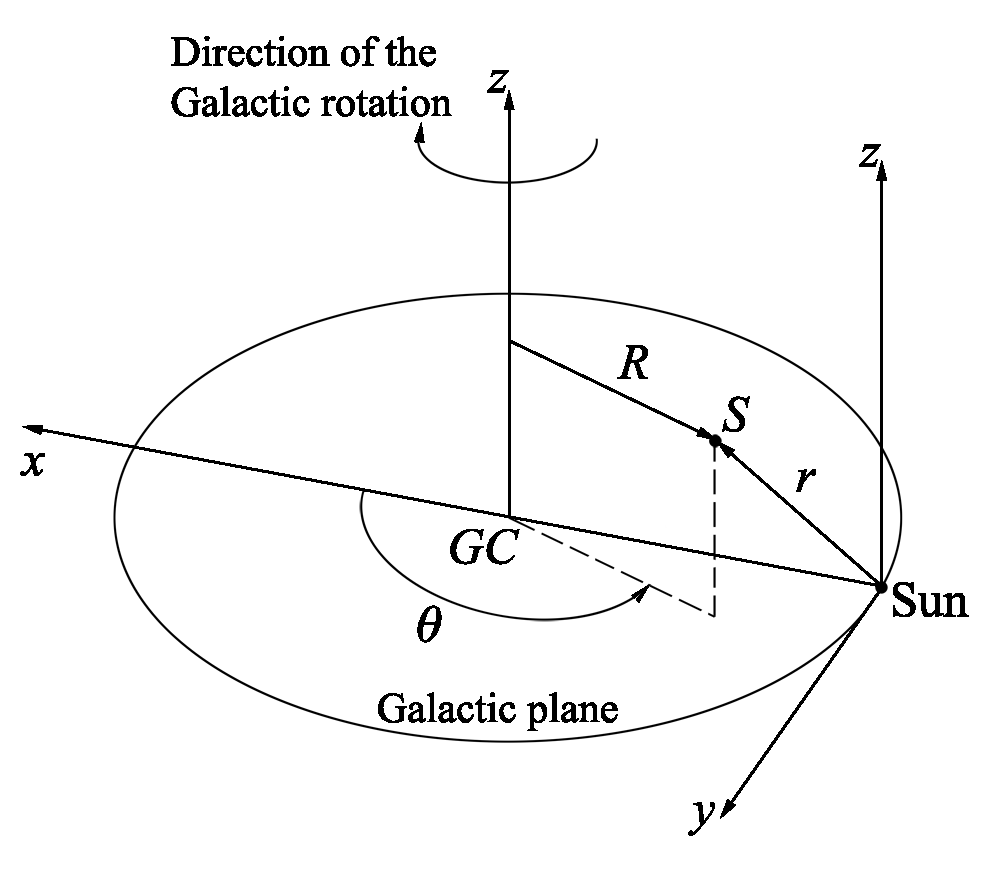}}
   \caption{The local rectangular Galactic ($x, y, z$) and cylindrical Galactocentric ($R, \theta, z$) coordinate systems. $S$ is an arbitrary star with heliocentric distance $r$.}
\label{fig:GCS}
\end{figure}

It is convenient to use the Cartesian coordinate system ($x, y, z$) for the basic kinematic equations of various models. The origin of the system coincides with the Solar System barycenter. The $x$ axis points to the Galactic center, the $y$ axis coincides with the direction of the Galactic rotation, while $z$ is perpendicular to the Galactic plane and complements the right-handed Cartesian coordinate system (see Fig. \ref{fig:GCS}). The given system is called the local (heliocentric) rectangular Galactic coordinate system.

The transformation of the positions and velocity components from the spherical coordinate system $(l, b, r, \mu_l, \mu_b, v_r)$ to the rectangular Cartesian one $(x, y, z, V_x, V_y, V_z)$ is
\begin{align}\label{sph2car}
  & x = r \cos b \cos l ,\nonumber \\
  & y = r \cos b \sin l ,\nonumber \\
  & z = r \sin b ,\\
  & V_x = V_r \cos l \cos b - V_l \sin l - V_b \cos l \sin b ,\nonumber \\
  & V_y = V_r \sin l \cos b + V_l \cos l - V_b \sin l \sin b  \nonumber \\
  & V_z = V_r \sin b + V_b \cos b ,\nonumber
\end{align}
where $V_r$ is the radial velocity, $V_l = k\,r\,\mu_l\,\cos b, V_b = k\,r\,\mu_b\,$ are the components of stellar proper motion, and $k$ = 4.74057 is the transformation factor from \myr to \kmskpc. The distance $r$ is calculated from the \emph{Gaia}~DR3 parallax as $1/\varpi$. Formulas (\ref{sph2car}) allow us to obtain the velocity field of the stars relative to the Sun in the rectangular coordinate system.

We calculated the kinematic parameters using both uncorrected \emph{Gaia} DR3 data and also considering the zero point correction for parallax and proper motion \citep{Lindegren2021,Cantat-Gaudin2021}. It turns out that for points with a heliocentric distance less than 5 kpc, the behaviour of the kinematic parameters does not change practically, except for a small "compression" in the direction of the Sun, which does not affect the behaviour of the parameters in any way. Some differences are possible at large distances from the Sun, more than 5-6 kpc, where the errors in estimating the astrometric parameters are large. In this work we have chosen to use uncorrected \emph{Gaia} DR3 data, since the main goal of the work is to demonstrate the capabilities of the presented method on \emph{Gaia} DR3 data, which will be corrected and improved several times in systematic and random accuracy in the next data releases.

We calculate the velocity components of each star that determine the velocity field relative to the motionless Galactocentric coordinate system as:
\begin{align}
    & V_{x,gal} = V_x + V_{x,\odot}, \nonumber\\
    & V_{y,gal} = V_y + V_{y,\odot},          \\
    & V_{z,gal} = V_z + V_{z,\odot}, \nonumber
\end{align}
where $V_x, V_y, V_z$ are the velocity components of each star relative to the Solar system barycenter and the values of the solar velocity components $(V_x, V_y, V_z)_\odot = (9.5, 250.7, 8.56)$  \kms are relative to the Galactic center \citep{Reid2020,Gravity2021,Drimmel2018}.

We estimate the mean 3D velocity distribution by sampling our stellar catalogue with cubic cells of $100$pc$\times$100pc$\times$100pc. In each cell, the mean  positions $(x, y, z)$ and spatial velocities ($V_x, V_y, V_z$) are derived by averaging the positions and velocities of the stars. These mean values are also called ''fictitious stars'' in the following sections.
The total number of cells (fictitious stars) is about 270 thousand. To perform the averaging, a minimum number of 3 stars is required.

\begin{figure*}
\begin{minipage}[h]{0.49\linewidth}
\includegraphics [width = 88mm] {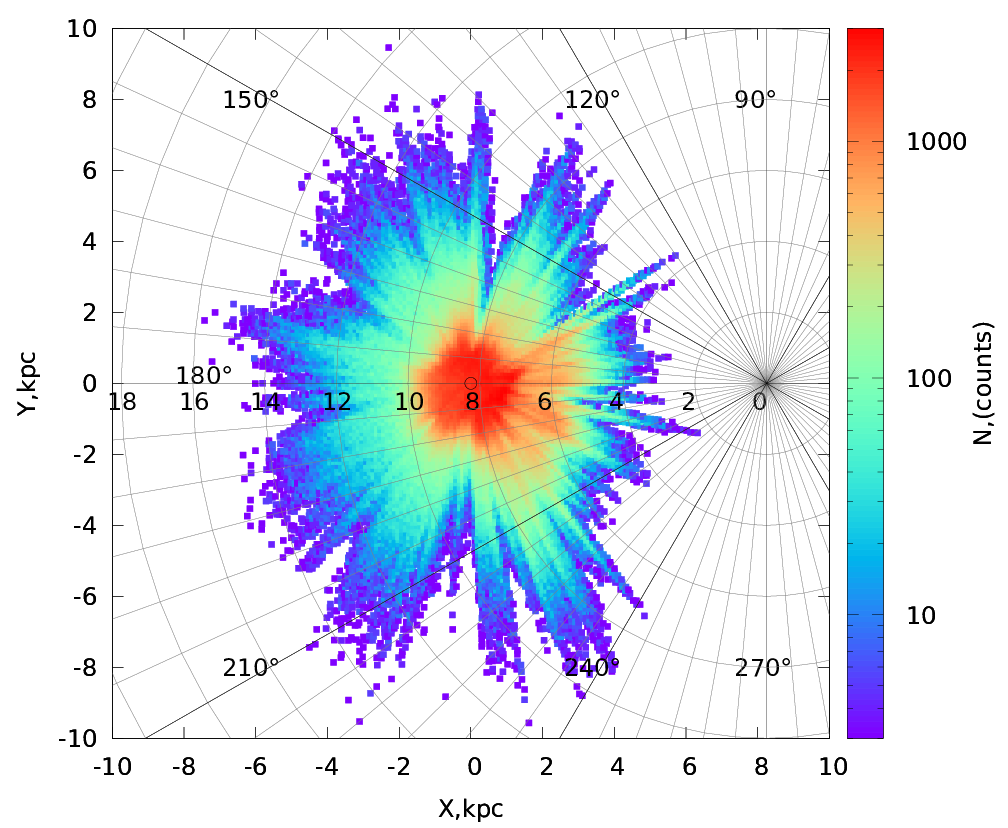}
\end{minipage}
\hfill
\begin{minipage}[h]{0.49\linewidth}
\includegraphics [width = 88mm] {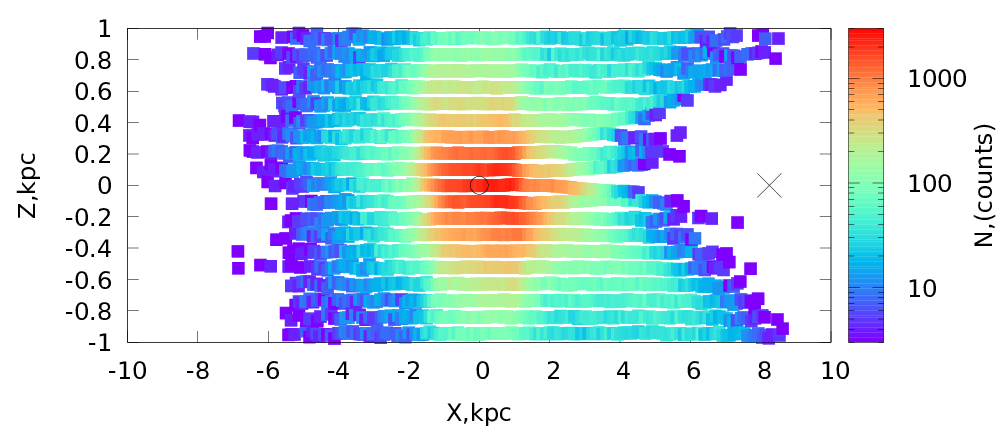}
\includegraphics [width = 88mm] {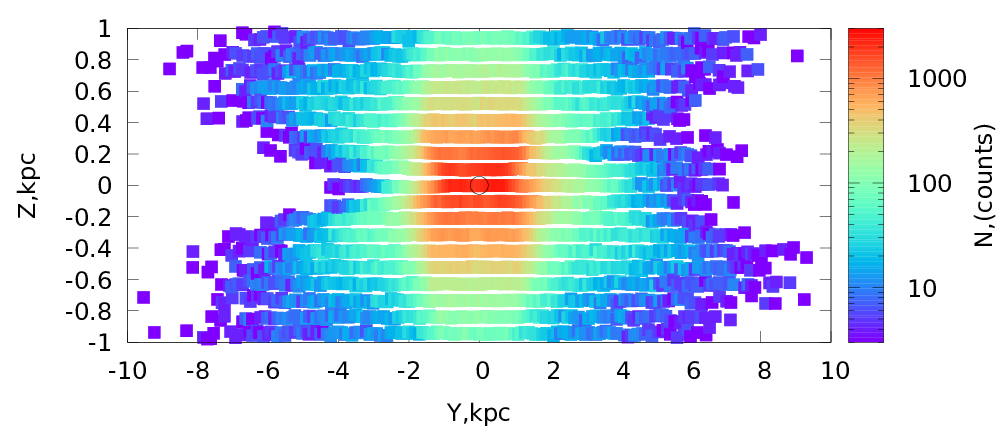}
\end{minipage}
\caption{Galactic maps of one-cell layer in Cartesian coordinates color coded according to the stellar count in the cell. The {\it left} shows the maps in Cartesian coordinates (x,y) on the galactic plane. The {\it right top} and {\it bottom} show the maps in Cartesian coordinates (x,z) and (y,z) in the y=0 and x=0 planes, respectively. The distance to the Sun $R_{\odot}=8.249$~kpc is marked by the point ${\odot}$ here and in all the following figures.}
\label{fig:density_stars}
\end{figure*}

Fig. \ref{fig:density_stars} shows the stellar density in (x,y), (y,z) and (x,z) planes for the cell size of $100$pc$\times$100pc$\times$100pc in the Cartesian heliocentric system (Fig. \ref{fig:GCS}).   
Fig. \ref{fig:density_stars} clearly shows the incompleteness due to the strong interstellar extinction towards low Galactic latitudes. In particular, the lack of stars close to the Galactic plane is apparent in the inner disc ($X>4$ kpc) and towards $Y>$4-5 kpc from the Sun.

The conversion from the Cartesian heliocentric system to the Galactocentric cylindrical coordinate system also requires knowledge of the distance of the Sun to the Galactic center \mbox{$R_\odot = 8.249$}~kpc, and height of the Sun from the Galactic plane of $Z_\odot = 0.0208$~kpc \citep{ Bennett2019,Gravity2021}. We have applied the transformation procedure to the Galactic cylindrical coordinate system to each cell (fictitious star):
\begin{align}
    & R = \sqrt{(R_\odot-x)^2 + y^2}, \nonumber\\
    & \theta = 180^\circ-\arctan{\frac{y}{R_\odot-x}},          \\
    & Z = z+Z_\odot, \nonumber \\
     \label{eq:gal_vel}
    & V_R = V_{x,gal} \cos \theta + V_{y,gal} \sin \theta, \nonumber\\
    & V_\theta = V_{y,gal} \cos \theta - V_{x,gal} \sin \theta,\\
    & V_Z = V_{z,gal}. \nonumber    
\end{align}


\section{Methods for determination of kinematic parameters}
\label{sec:methods}

The Ogorodnikov-Milne (O--M) model is based on Helmholtz's theorem for stellar systems: within a small region surrounding any point in a stellar system, the velocities of the centroids can be modelled as the superposition of a rigid translation and rotation velocity plus a deformation component velocity of the region \citep{Ogorodnikov1965}. \cite{Ogorodnikov1932} proposed this model and \cite{Milne1935} generalised it to the three-dimensional case. The stellar velocity field, which fills a given spatial region, is treated as a continuous deformable medium, which allows us to study stellar motion using the methods of continuum mechanics.

In the Cartesian Galactic coordinate system ($x_k, k=1,2,3$), the general form of the expansion of the velocity field ${\textbf V}({\textbf r})$ in the vicinity of the centroid located at ${\bf r}_0$ is:
\begin{multline}
 {\bf V}({\bf r}) =  {\bf V}({\bf r}_0) + \left( \frac{\partial{\bf V}}{\partial x_k}\right)_0 d{x_k}  + \\
 + \frac{1}{2!}\left( \frac{\partial^2 {\bf V}}{\partial x_k \partial x_m}\right)_0 d{x_k}  d{x_m} + ..., (k, m =1, 2, 3). 
\label{eq:velocity_field}
\end{multline}
where "0" denotes the derivatives at $x_{k0} =(x_0,y_0,z_0)$.
In order to analyze the stellar velocities $ V_i$ inside the sphere with radius  $r\ll R_0$ (distance to the Galactic Centre) in the O--M model, we can restrict the expansion of the series to the first order terms:
\begin{multline}
  V_i({\bf r}_0 + \partial {\bf r}) = V_i({\bf r}_0) 
 + \frac{1}{2} \left(\frac{\partial V_i}{\partial x_k} 
 - \frac{\partial V_k}{\partial x_i} \right)_0 d{x_k} + \\
 + \frac{1}{2} \left(\frac{\partial V_i}{\partial x_k} 
 + \frac{\partial V_k}{\partial x_i} \right)_0 d{x_k} 
 = V_i({\bf r}_0) + \omega_{ik} d{x_k} + V_{ik} d{x_k}.
  \label{eq:velocity_field_part}
\end{multline}
Here
\begin{align}
\omega_{ik} = \frac{1}{2} \left(\frac{\partial V_i}{\partial x_k} 
 - \frac{\partial V_k}{\partial x_i} \right)_0 = -\omega_{ki}
\end{align}
and 
\begin{align}
V_{ik} = \frac{1}{2} \left(\frac{\partial V_i}{\partial x_k} 
 + \frac{\partial V_k}{\partial x_i} \right)_0 = V_{ki}
\end{align}
are components of antisymmetric $M^-$ and symmetric $M^+$ tensors respectively:
\begin{align}
M^{\pm} = \frac{1}{2} \left(\frac{\partial V_i}{\partial x_k} 
 \pm \frac{\partial V_k}{\partial x_i} \right)_0
 \label{eq:matrix_as}
\end{align}
and $i, k = 1, 2, 3$.
The antisymmetric tensor $M^-_{ik}=\omega_{ik}$ is also called the local rotation tensor, since it is equivalent to the axial vector ${\boldsymbol \omega}(\omega_1,\omega_2,\omega_3)$, where $\omega_1 = M^-_{32}$, $\omega_2 = M^-_{13}$, $\omega_3 = M^-_{21}$. The second-rank symmetric tensor ${\sf V}_{ik}$ is called the local deformation velocity tensor.
Taking into account (\ref{eq:matrix_as}) we can rewrite the expression (\ref{eq:velocity_field_part}) in the following form:
\begin{equation}
 V_i({\bf r}_0 + d{\bf r}) =  V_i({\bf r}_0) + M^- d{x_k} + M^+ d{x_k}.
\end{equation}
The elements of the matrices $M^-$ and $M^+$ are partial derivatives with respect to the coordinates of the projections ($V_1, V_2, V_3$) of the velocity vector onto the rectangular coordinate axes. These elements are referred to as the kinematic parameters. The relations between the kinematic parameters of the O--M model and the partial derivatives of the first order velocity components with respect to Galactocentric coordinates ($R, \theta, Z$) are as follows:
\begin{align}
& M^+_{11} = \frac{\partial V_R}{\partial R}\, \label{eq:M11} \\
& (\omega_3 + M^+_{12}) = \frac{\partial V_\theta}{\partial R} = -\frac{\partial V_{\rm rot}}{\partial R} \label{eq:dVtheta_dR_om} \, ,\\
& (\omega_2 - M^+_{13}) = \frac{\partial V_Z}{\partial R}    \label{eq:dVz_dR_om}   \, ,              \\
& (\omega_3 - M^+_{12}) = \frac{V_\theta}{R} - \frac{1}{R} \frac{\partial V_R}{\partial\theta}, \label{eq:W3_M12p}\\
& M^+_{22} = \frac{V_R}{R} + \frac{1}{R} \frac{\partial V_\theta}{\partial\theta} = \frac{V_R}{R} - \frac{1}{R} \frac{\partial V_{\rm rot}}{\partial\theta} \, , \\
& (\omega_1 + M^+_{23}) = -\frac{1}{R} \frac{\partial V_Z}{\partial\theta}  \label{warp} \, ,\\
& (\omega_2 + M^+_{13}) = -\frac{\partial V_R}{\partial Z}  \label{eq:dVR_dz_om} \, ,\\
& (\omega_1 - M^+_{23}) = \frac{\partial V_\theta}{\partial Z} = -\frac{\partial V_{\rm rot}}{\partial Z} \label{eq:dVtheta_dz_om} \, ,\\
& M^+_{33} = \frac{\partial V_Z}{\partial Z}\, .  \label{eq:dVz_dz}
\end{align}
where $V_\theta$ has the opposite direction with respect to the circular rotation velocity of the Galaxy $V_{\rm rot}$.

We can also determine the Oort constants $A$, $B$, $C$ and $K$ which measure the azimuthal shear, the vorticity, the radial shear, and the divergence of the velocity field in the Galactic plane respectively. The Oort constants in Galactocentric cylindrical coordinates $(R, \theta)$ are given by \citep{Chandrasekhar1942,Ogorodnikov1965}:
\begin{align}
& A = \frac{1}{2} \left(\frac{\partial V_x}{\partial y} + \frac{\partial V_y}{\partial x} \right) = \frac{1}{2}\Big(\frac{\partial V_{\theta}}{\partial R} -\frac{V_{\theta}}{R} + \frac{1}{R} \frac{\partial V_R}{\partial \theta} \Big), \label{eq:Oort_A}\\
& B =  \frac{1}{2} \left(\frac{\partial V_y}{\partial x} - \frac{\partial V_x}{\partial y} \right) = \frac{1}{2}\Big(\frac{\partial V_{\theta}}{\partial R} +\frac{V_{\theta}}{R} - \frac{1}{R} \frac{\partial V_R}{\partial \theta} \Big), \label{eq:Oort_B}\\
& C = \frac{1}{2} \left(\frac{\partial V_x}{\partial x} - \frac{\partial V_y}{\partial y} \right) =\frac{1}{2}\Big(\frac{\partial V_R}{\partial R} -\frac{V_R}{R} -  \frac{1}{R} \frac{\partial V_{\theta}}{\partial \theta} \Big), \label{eq:Oort_C}\\
& K =\frac{1}{2} \left(\frac{\partial V_x}{\partial x} + \frac{\partial V_y}{\partial y} \right)=\frac{1}{2}\Big(\frac{\partial V_R}{\partial R} +\frac{V_R}{R} +  \frac{1}{R} \frac{\partial V_{\theta}}{\partial \theta} \Big).\label{eq:Oort_K}
\end{align}
Formulas (\ref{eq:M11}) -- (\ref{eq:dVz_dz}) do not allow us to determine the parameters $\frac{\partial V_R}{\partial\theta}$ and $\frac{\partial V_\theta}{\partial\theta}$ from combinations of the O--M model parameters. It is therefore necessary to introduce some reasonable assumptions about these derivatives. 
 
In particular, the combination of the parameters $\omega_3$ and $M^+_{12}$ from formula (\ref{eq:W3_M12p}), which is derived within the O--M model, gives us the following expression:
\begin{equation}
    V_{\rm rot} = -V_\theta = (M^+_{12} - \omega_3)R - \frac{\partial V_R}{\partial\theta}. \label{eq:Vrot_dVR}
\end{equation}
Since in the Oort-Lindblad model it was assumed that the stellar system under consideration is axisymmetric, i.e. $\partial V_R/\partial\theta = 0$ and $\partial V_{\theta}/\partial\theta = 0$, we can write (\ref{eq:Vrot_dVR}) via the Oort constants $A$ and $B$:
\begin{equation}   \label{eq:Vrot}
    V_{\rm rot} = (A-B)R.
\end{equation}
This is a commonly used formula for the determination of $V_{\rm rot}$ \citep{ Milne1935, Chandrasekhar1942, Ogorodnikov1965, Clube1972, duMont1977, Miyamoto1993, Miyamoto1998, Vityazev2005, Vityazev2009, Bobylev2011, Bovy2017}.

The O-M model has been used by many authors to find the kinematic parameters of the Galaxy at one point - near the Sun. Since the \emph{Gaia} DR3 catalogue provides a set of spatial positions and velocities of the stars that extend well beyond the solar neighborhood, we can perform similar kinematic studies at any point of the Galaxy as shown in \citet{Fedorov2021, Fedorov2023}.

First of all, we create a rectangular grid specified in the Galactic plane where $z=0$. The grid nodes are separated from each other along the coordinates $x$ and $y$ by a distance of 100 pc in the range $-10$\,kpc and $+10$\,kpc for $X$ and $Y$ axes of the heliocentric coordinate system. The total number of nodes in the Galactic plane is 40 thousand. For each node of the grid, we select a spherical region with a radius of 1.0 kpc. Each sphere around a given node contains a set of $n=1, 2, 3....N$ cells (fictitious stars) that are located at distances not exceeding 1 kpc, i.e. each $n$-th fictitious star satisfies the condition:
\begin{equation}
\sqrt{(x_n-x)^2+(y_n-y)^2+(z_n)^2}<1~\textrm{kpc} \label{eq:r_1}.
\end{equation}

The optimal size of the spheres was chosen after numerous calculations using different radii. As a result, we obtained a radius of 1 kpc, which is a compromise between a large number (accuracy) of fictitious stars and the high resolution (detail) of the parameter maps. When the radius of the spheres is small, the obtained kinematic parameters are sensitive to local variations in the stellar velocity field. Since we want to study the global behaviour of the kinematic parameters, we use a radius of 1 kpc, which is optimal for these scales. 

\begin{figure}
\includegraphics [width = 90mm] {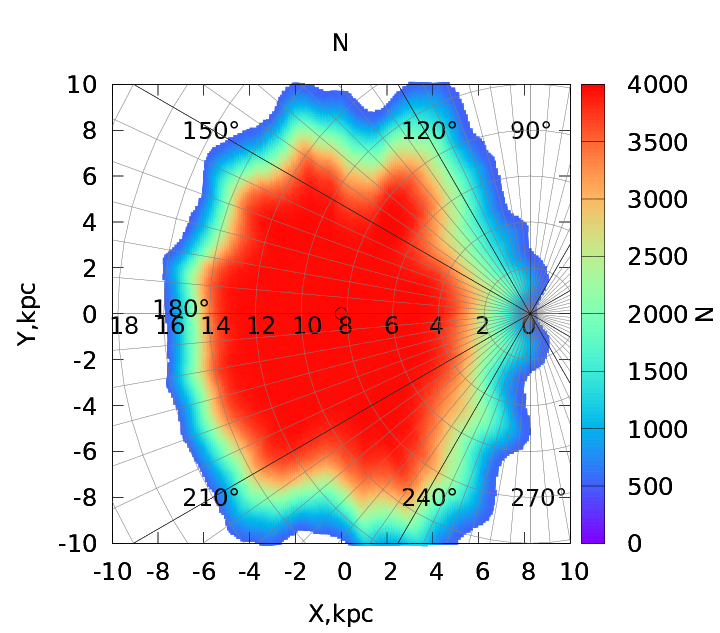}
\caption{The count of the cells (fictitious stars) in the selected spherical regions with radius of 1.0 kpc that were used for calculating the kinematic parameters.}
\label{fig:Nbin_stars}
\end{figure}

The kinematic parameters were calculated for each node with $x$ and $y$ coordinates and located at a distance of 100pc$\times$100pc in the Galactic plane. Thus,  the overlap of our spherical samples provides for the continuity and differentiability of the velocity field, which makes it possible to study global kinematic parameters instead of studying them as discrete values at local points. Only spheres containing $N \geq500$ cells (fictitious stars) have been used in this investigation (Fig. \ref{fig:Nbin_stars}). However, we trust the results of calculations for those points on the map for which the number of cells (fictitious stars) is more than 2000 which corresponds to a heliocentric distance of 5-6 kpc (Fig. \ref{fig:Nbin_stars}).

We thus form 28139 spherical regions (centroids) with coordinates $x$ and $y$ in the Galactic plane ($Z=0$), each containing $500<N<3981$ fictitious stars. For each spherical region we compute the cylindrical Galactocentric velocity components and their first and second order partial derivatives using the position and velocity of the fictitious stars from the solution of the equations by means of the least squares method (LSM):
\begin{equation}
\begin{split}
&V_{R}(R_n,\theta_n,Z_n) = \Big(V_R\Big)_{xy} +\Big(\frac{\partial V_R}{\partial R}\Big)_{xy} (R_n-R_{xy})+ \\
&+\Big(\frac{\partial V_R}{\partial \theta}\Big)_{xy} (\theta_n-\theta_{xy}) + \Big(\frac{\partial V_R}{\partial Z}\Big)_{xy}  Z_n+\\
&+\frac{1}{2}\Big(\frac{\partial^2 V_R}{\partial R^2}\Big)_{xy}(R_n-R_{xy})^2 + \Big(\frac{\partial^2 V_R}{\partial R \partial \theta}\Big)_{xy} (R_n-R_{xy})(\theta_n-\theta_{xy}) +\\
&+\Big(\frac{\partial^2 V_R}{\partial R \partial Z}\Big)_{xy} (R_n-R_{xy})Z_n + \frac{1}{2}\Big(\frac{\partial^2 V_R}{\partial \theta^2 }\Big)_{xy} (\theta_n-\theta_{xy})^2 +\\
&+\Big(\frac{\partial^2 V_R}{\partial \theta \partial Z}\Big)_{xy} (\theta_n-\theta_{xy}) Z_n + \frac{1}{2}\Big(\frac{\partial^2 V_R}{\partial Z^2 }\Big)_{xy} Z_n^2,
\label{eq:derivativesVR} 
\end{split}
\end{equation}

\begin{equation}
\begin{split}
&V_{\theta}(R_n,\theta_n,Z_n)= \Big(V_\theta\Big)_{xy}+\Big(\frac{\partial V_\theta}{\partial R}\Big)_{xy} (R_n-R_{xy}) +\\ 
& +\Big(\frac{\partial V_\theta}{\partial \theta}\Big)_{xy}(\theta_n-\theta_{xy}) + \Big(\frac{\partial V_\theta}{\partial Z}\Big)_{xy} Z_n+\\
&+\frac{1}{2}\Big(\frac{\partial^2 V_\theta}{\partial R^2}\Big)_{xy}(R_n-R_{xy})^2
+\Big(\frac{\partial^2 V_\theta}{\partial R \partial \theta}\Big)_{xy} (R_n-R_{xy})(\theta_n-\theta_{xy}) +\\
&+\Big(\frac{\partial^2 V_\theta}{\partial R \partial Z}\Big)_{xy}(R_n-R_{xy}) Z_n + \frac{1}{2}\Big(\frac{\partial^2 V_\theta}{\partial \theta^2 }\Big)_{xy} (\theta_n-\theta_{xy})^2 +\\
&+\Big(\frac{\partial^2 V_\theta}{\partial \theta \partial Z}\Big)_{xy}(\theta_n-\theta_{xy}) Z_n+\frac{1}{2}\Big(\frac{\partial^2 V_\theta}{\partial Z^2 }\Big)_{xy} Z_n^2,
\label{eq:derivativesVth}
\end{split}
\end{equation}

\begin{equation}
\begin{split}
& V_{Z}(R_n,\theta_n,Z_n) = \Big(V_Z\Big)_{xy} +\Big(\frac{\partial V_Z}{\partial R}\Big)_{xy} (R_n-R_{xy})+\\
&+\Big(\frac{\partial V_Z}{\partial \theta}\Big)_{xy} (\theta_n-\theta_{xy}) +\Big(\frac{\partial V_Z}{\partial Z}\Big)_{xy} Z_n +\\ 
&+\frac{1}{2}\Big(\frac{\partial^2 V_Z}{\partial R^2}\Big)_{xy}(R_n-R_{xy})^2 + \Big(\frac{\partial^2 V_Z}{\partial R \partial \theta}\Big)_{xy}(R_n-R_{xy})(\theta_n-\theta_{xy}) + \\
&+\Big(\frac{\partial^2 V_Z}{\partial R \partial Z}\Big)_{xy} (R_n-R_{xy}) Z_n +\frac{1}{2}\Big( \frac{\partial^2 V_Z}{\partial \theta^2 }\Big)_{xy} (\theta_n-\theta_{xy})^2 +\\
& +\Big(\frac{\partial^2 V_Z}{\partial \theta \partial Z}\Big)_{xy}(\theta_n-\theta_{xy}) Z_n +
\frac{1}{2}\Big(\frac{\partial^2 V_Z}{\partial Z^2 }\Big)_{xy} Z_n^2.
\label{eq:derivativesVz}
\end{split}
\end{equation}
These formulas are obtained by expanding the cylindrical Galactocentric velocity components $V_{R}(R,\theta,Z)$, $V_{\theta}(R,\theta,Z)$, $V_{Z}(R,\theta,Z)$ into a Taylor series limited by the second order terms of all fictitious stars with coordinates $R_n, \theta_n$ and $Z_n$ satisfying condition \ref{eq:r_1}. The expansions were made in the vicinity of the points with coordinates $(R_{xy},\theta_{xy},0)$. These points coincide with the nodes of the rectangular grid $(x,y)$. Condition equations (\ref{eq:derivativesVR}), (\ref{eq:derivativesVth}), (\ref{eq:derivativesVz}) were formed for each $n$-th fictitious star, and the system of equations was solved by least squares in 30 unknowns: three velocity components $V_R, V_\theta, V_Z$,  their nine partial derivatives of the first order and eighteen partial derivatives of the second order with respect to coordinates for each point with coordinate $x,y$. The system of equations was compiled and solved independently for each spherical region (centroid) with $x$ and $y$ coordinates. Thus, in the Galactic plane we obtained 28139 solutions of the equations system (\ref{eq:derivativesVR}), (\ref{eq:derivativesVth}), (\ref{eq:derivativesVz}) and each solution containing 30 kinematic parameter estimates and their errors.
The found kinematic parameters describe the stellar velocity field at any given point with  $x$ and $y$ coordinates. This approach allows us to represent the behaviour of the calculated kinematic parameters from Galactic coordinates $(x,y)$ or $(R,\theta)$. 

We have calculated the kinematic parameters in two ways: the first method is the use of "fictitious stars" (as presented in this paper) and the second method is the use of individual stars in equations(\ref{eq:derivativesVR}), (\ref{eq:derivativesVth}), (\ref{eq:derivativesVz}). 
There are practically no differences between the results of the two methods, but the differences in the speed of calculations are significant. For this reason, we show the results of the faster computational method (using "fictitious stars"), which is particularly important for future \emph{Gaia} releases that will contain more data.

\section{Analysis of the results}
\label{sec:analysis}

In this Section we present the distribution of the kinematic parameters estimated from Eqs.\ \ref{eq:derivativesVR}-\ref{eq:derivativesVz} in the form of colour maps. 
The kinematic parameters derived by the O--M model are in good agreement over the entire coordinate range with the stellar velocity field estimated from a Taylor series limited by the first order in the cylindrical Galactocentric coordinate system, as also found by \citet{Fedorov2023}. Furthermore, our approach provides additional kinematic parameters, including those that cannot be determined with the O--M model. On the colour maps, the coordinate plane ($x, y$) coincides with the Galactic plane. The kinematic parameters in each point of this plane are calculated in the Galactocentric cylindrical coordinate system $R$, $\theta$ and $Z=0$. The Sun is represented as $\odot$ with coordinates ($x, y$) = (0, 0) and ($R_{\odot}, \theta$) = $(8.249 \text{ kpc}, 180^\circ)$ .

The mean value of the random errors of the kinematic parameters is 1.5-2 \kmskpc up to heliocentric distances of about 8 kpc for the first order derivatives and 5-6 kpc for those of the second order; there is no correlation between the behaviour of the parameters and their errors. This gives us reason to consider these parameters as real.
The actual distribution of the random errors of the kinematic parameters are shown in Appendix \ref{sec:app1} (Fig. \ref{fig:eV}, \ref{fig:edV1}, \ref{fig:edV2} and \ref{fig:edV2mix}). 

\begin{figure*}
\includegraphics[width = 88mm]{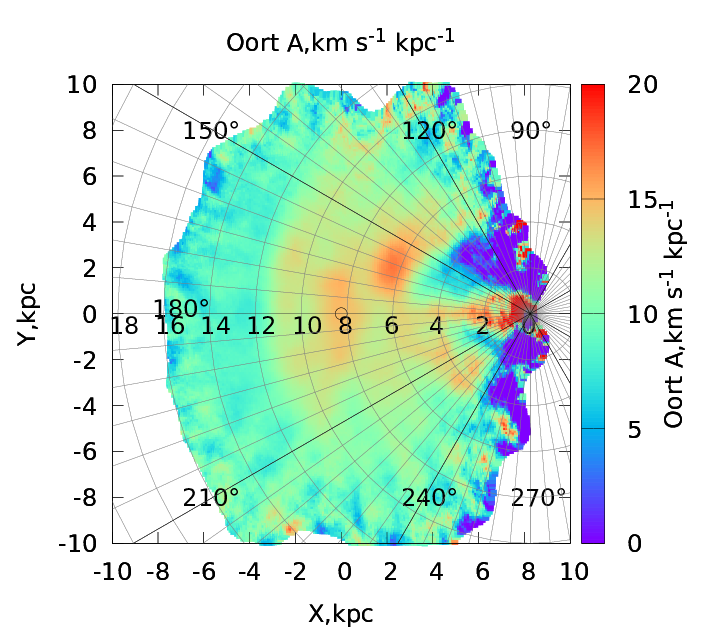}
\includegraphics[width = 88mm]{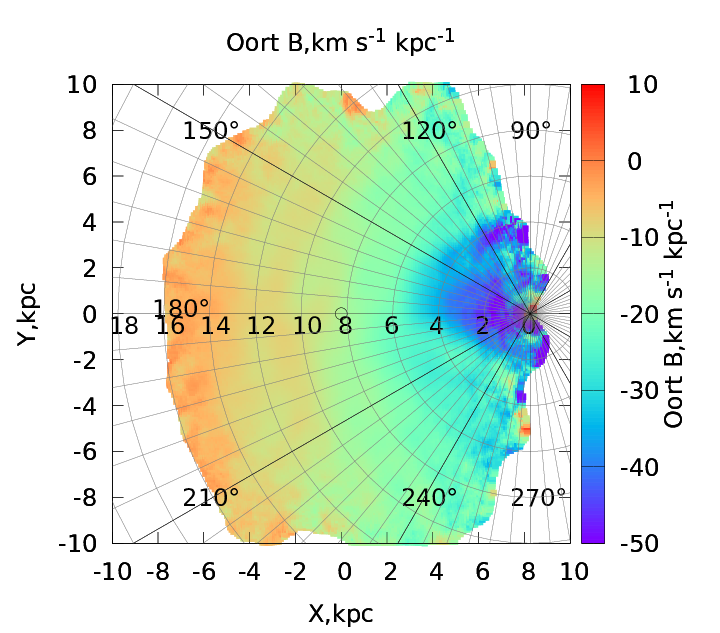}
\includegraphics[width = 88mm]{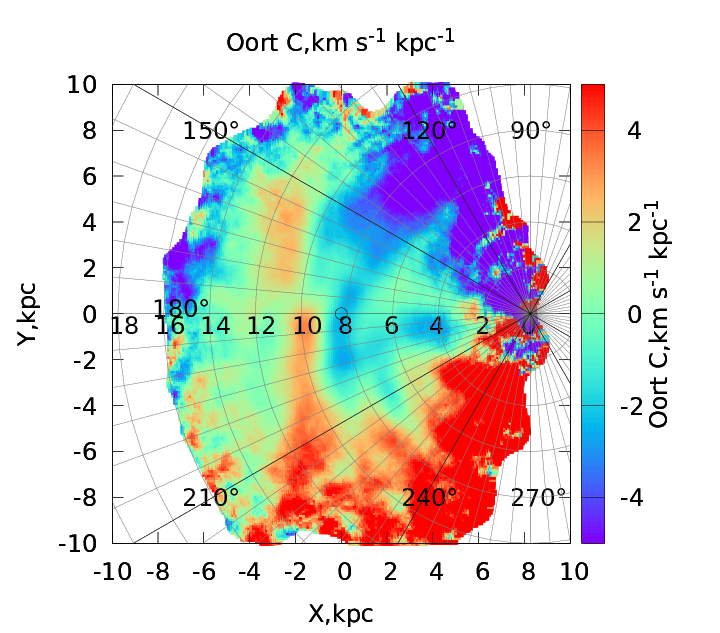}
\includegraphics[width = 88mm]{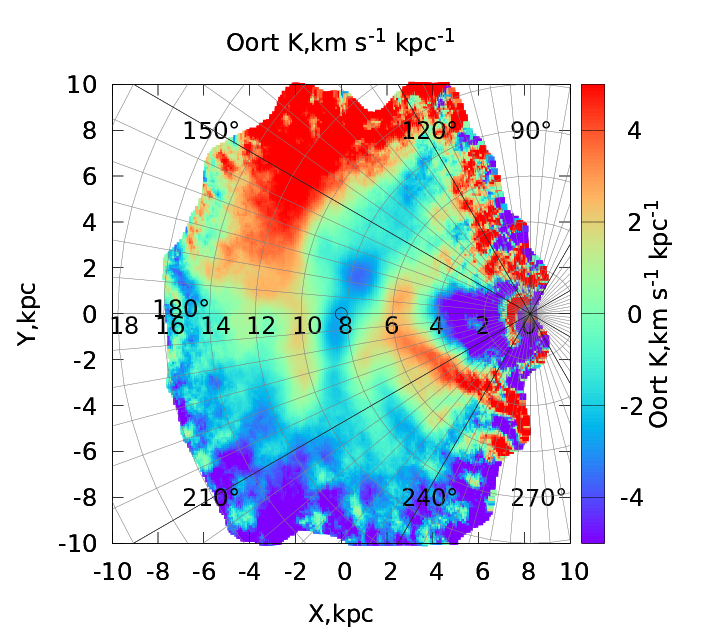}
\caption{Oort constants $A$, $B$, $C$ and $K$ calculated from the equations (\ref{eq:Oort_A}), (\ref{eq:Oort_B}), (\ref{eq:Oort_C}) and (\ref{eq:Oort_K}) by data solution from the equation (\ref{eq:derivativesVR}) and (\ref{eq:derivativesVth}) as functions of the Galactic coordinates.}
\label{fig:Oort_constant}
\end{figure*}

\subsection{ Oort constants $A$, $B$, $C$ and $K$}

In figure \ref{fig:Oort_constant} we show the values of the Oort constants $A$, $B$, $C$ and $K$ calculated from the equations (\ref{eq:Oort_A}), (\ref{eq:Oort_B}), (\ref{eq:Oort_C}) and (\ref{eq:Oort_K}) in the whole Galactic plane for different Galactocentric distances and angles, and their change in different directions. Our estimations of the Oort constants are listed in Tab. \ref{tab:Oort_constant} and are in good agreement with those made by \citet{Siebert2011}, \citet{Bobylev2021}, \citet{Bovy2017}, \citet{Vityazev2018}, \citet{Wang2021}. They calculated the Oort constants only at one point using stellar samples at distances of 0.5 or 1.5 kpc near the Sun. Similarly, a reasonable agreement with the values of these parameters in the Galactic plane was found in \citet{Nelson2022}.

It results from Figure \ref{fig:Oort_constant} that the Oort constants $C$ and $K$ deviate significantly from zero, reaching the values of \mbox{$\pm$(4-5) \kmskpc} as a function of the Galactic coordinates $R$ and $\theta$. The dependence of the Oort constants $A$ and $B$ on the distance $R$ and angle $\theta$ indicates a non-axisymmetric Galactic potential. It is also evident from figure \ref{fig:Oort_constant} that the Oort constants $A$, $C$ and $K$ show significant variations with $\theta$.

\subsection{Components of the stellar velocity field}

\begin{figure*}
\includegraphics[width = 58mm]{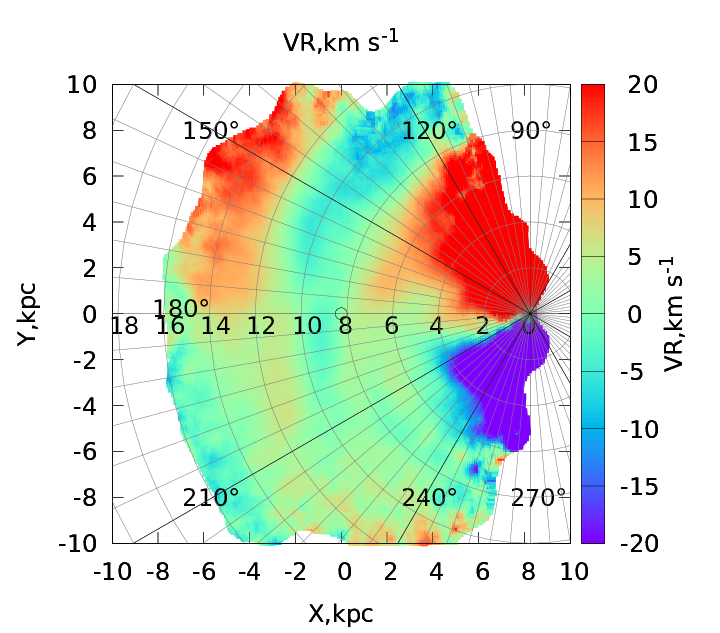}
\includegraphics[width = 58mm]{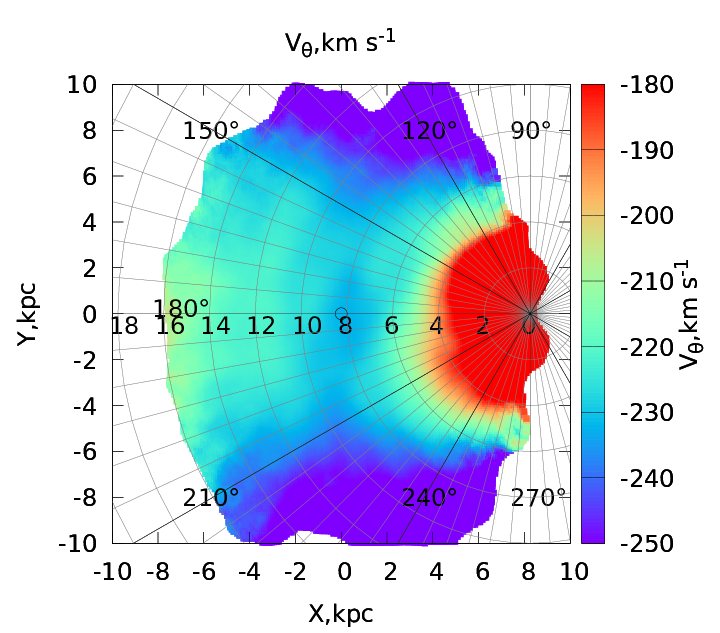}
\includegraphics[width = 58mm]{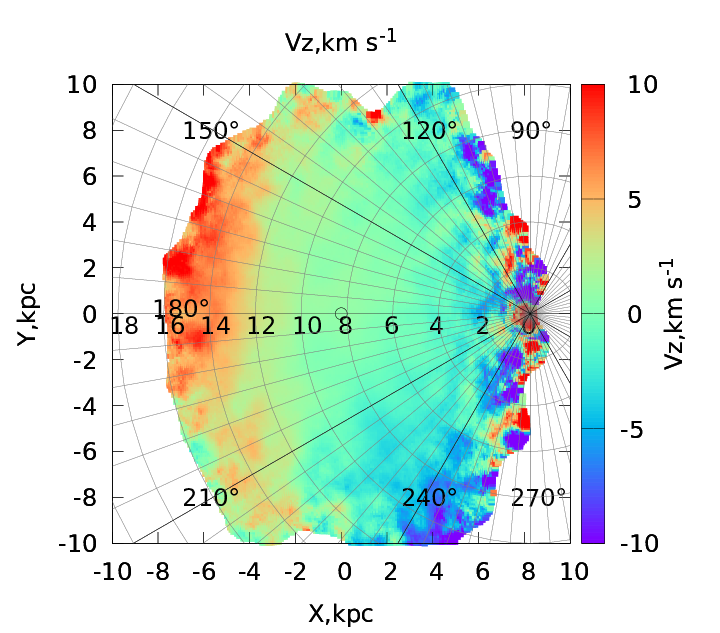}
\includegraphics[width = 58mm]{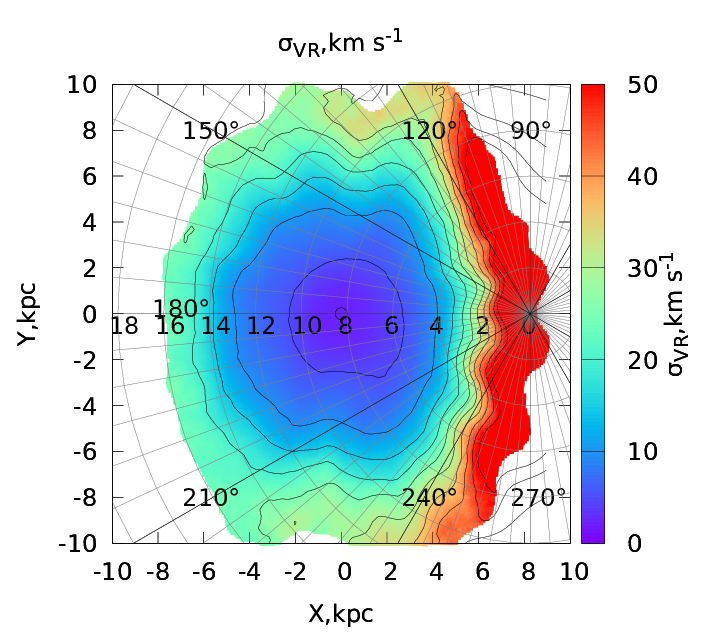}
\includegraphics[width = 58mm]{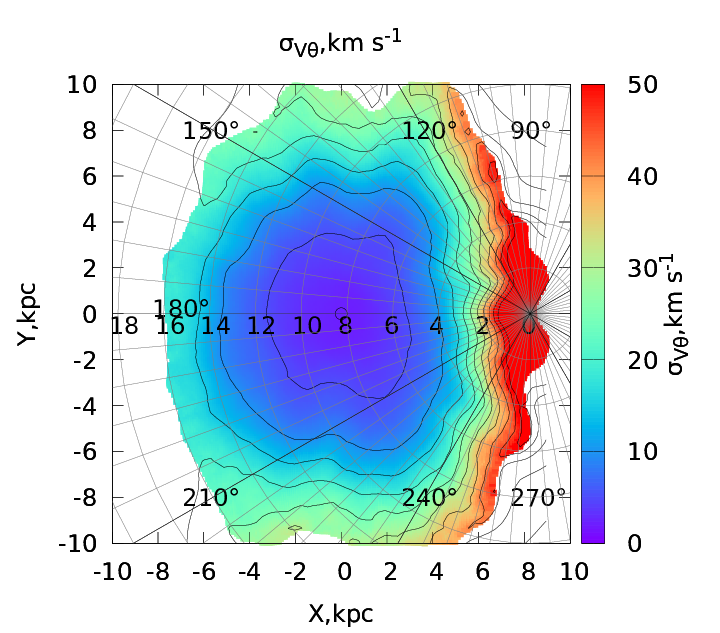}
\includegraphics[width = 58mm]{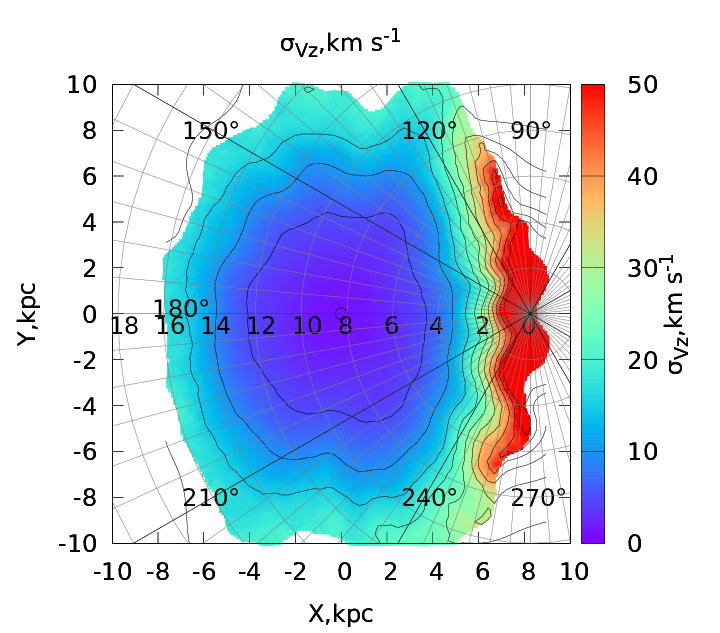}
\caption{{\it Top}: Components of the stellar velocity fields $V_R, V_\theta, V_Z$ calculated using the equations (\ref{eq:derivativesVR}), (\ref{eq:derivativesVth}) and (\ref{eq:derivativesVz}). {\it Bottom}:  Their standard deviation $\sigma_{ VR }, \sigma_{V\theta}, \sigma_{Vz}$ obtained by solving the equation system together with LSM.}
\label{fig:Velocity_residual}
\end{figure*}

Figure \ref{fig:Velocity_residual} shows the ordered ({\it top}) and the random ({\it bottom}) stellar velocity components $V_R, V_\theta, V_Z$ calculated using equations (\ref{eq:derivativesVR}), (\ref{eq:derivativesVth}) and (\ref{eq:derivativesVz}). The stellar velocity dispersions are estimated by the standard deviation $\sigma_{ V_R }, \sigma_{V_{\theta}}, \sigma_{V_Z}$ resulting from the LSM solution of  the  spherical regions with the radius of 1.0 kpc, as described in Section~\ref{sec:methods}. Our velocity distributions $(V_R, V_{\theta}, V_Z)$ as well as their standard deviations are in good agreement with previous kinematics studies \citep{Drimmel2022, Queiroz2021}. 

It is seen from the right panels of figure \ref{fig:Velocity_residual} that the values of the vertical stellar motions ${V_Z}$ and their standard deviation $\sigma_{V_Z}$ are the smallest from all the components. The vertical velocity component $V_Z$ practically does not change in the range of Galactocentric distances from 1 to 10 kpc and has a value close to zero. The vertical velocity increases with the Galactocentric radius for distances larger than 10 kpc  showing the kinematic signature of the Galactic warp and in good agreement with the results found by \citet{Poggio2018} and \citet{Katz2018}. 

In the middle panel of Fig. \ref{fig:Velocity_residual} we can see the azimuthal component of the Galaxy velocity obtained from solving the equation (\ref{eq:derivativesVth}). Note that $V_\theta$ has the opposite direction with respect to the Galaxy rotation velocity $V_{rot}$. The values of the Galactic rotation velocity differ slightly for centroids located in the range of coordinate angles $\theta$ between $150^{\circ}$ and $210^{\circ}$ and dependence from the Galactocentric distance. Thus, we can average the rotational Galaxy velocity in this range of $\theta$ angles, as shown in Fig. \ref{fig:Vrot} and Tab. \ref{tab:galaxy_curve_1} and \ref{tab:galaxy_curve_2}.
In other regions of the Galaxy whit $\theta <150^{\circ}$ or $\theta>210^{\circ}$ the velocity $V_{\theta}$ is not only a function of the Galactocentric distance R but also depends on the angle $\theta$. 
This behavior of the Galaxy's rotation velocity is caused by several factors. The first is a large error in the measurement of parallaxes at significant heliocentric distances. The second is a rapid decrease in the number of stars near the Galactic plane due to interstellar extinction. As a result, the azimuthal velocity $V_\theta$ can be biased.

In addition, the top-left panel of figure \ref{fig:Velocity_residual} evidences the symmetric inward and outward $V_R$ motion generated by the Galactic bar up to $R\la 6$ kpc. This kinematic signature was first revealed by \citet{Bovy2019}, who analysed 3D velocities derived from APOGEE DR16 and Gaia DR2 data. Our figure \ref{fig:Velocity_residual} shows an extended map, where the details of this kinematic feature are well consistent with \citet{Drimmel2022}.

Note that the values of the velocity components derived by  \citet{Drimmel2022} were obtained by {\it averaging} the velocity components of the stellar samples, while  \citet{Fedorov2023} fitted the velocity field with first order Taylor series. Neither of these methods takes into account the change in the absolute value of the velocity with the distance from the Galactic plane due to the vertical velocity gradient \citep{Vityazev2012,Velichko2020}.
Conversely, our velocity components $V_R (R_{xy},\theta_{xy},0)$, $V_\theta (R_{xy},\theta_{xy},0)$, $V_Z (R_{xy},\theta_{xy},0)$ and the corresponding velocity dispersions  result from an analytical approximation of the stellar velocity field to the Galactic plane ($Z = 0$), whose second order term, i.e. $d^2 V/ dZ^2$, is able to model the vertical velocity gradient.  
This is an important improvement, since the number of stars with heliocentric distances larger than 5-6 kpc decreases drastically towards the Galactic plane due to the strong extinction, so that our estimate of the velocity field at $Z=0$ is dominated by stars located above the Galactic plane.

\subsection{The radial gradient of the stellar velocity field}

\begin{figure*}
\includegraphics[width = 58mm]{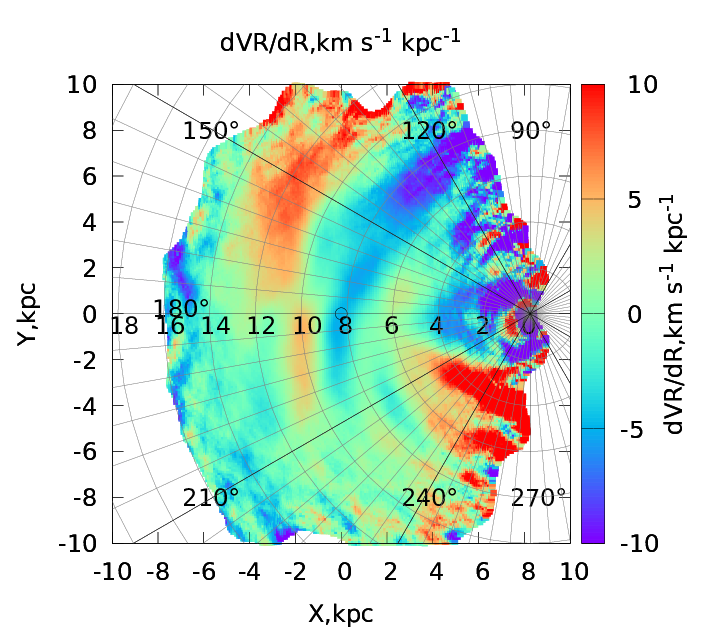}
\includegraphics[width = 58mm]{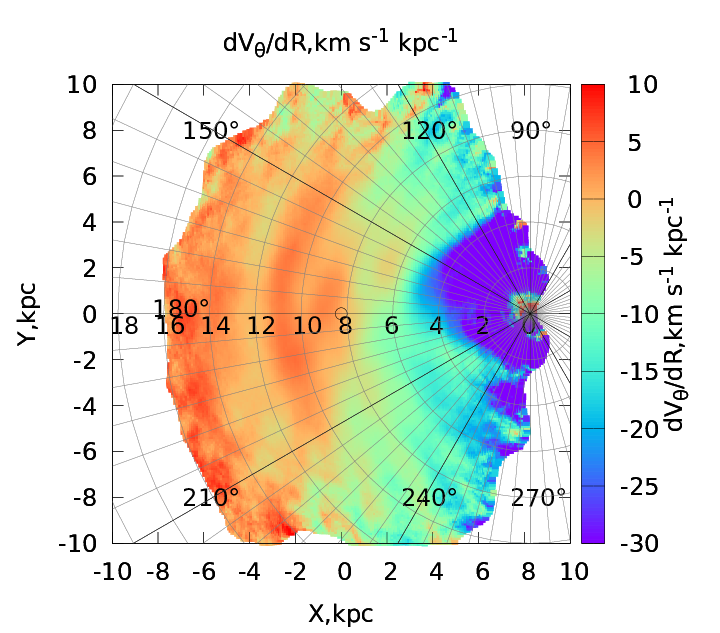}
\includegraphics[width = 58mm]{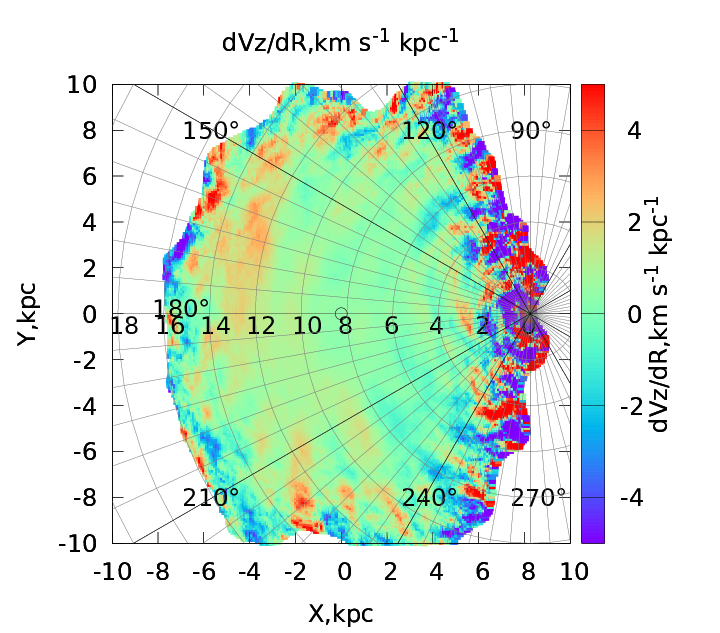}
\caption{The radial gradient of the velocity components (from left to right, respectively, $\frac{dV_R}{dR}, \frac{dV_{\theta}}{dR}, \frac{dV_Z}{dR})$ as a function of the Galactic coordinates.}
\label{fig:d_dR}
\end{figure*}

In Fig. \ref{fig:d_dR} we see the maps of the radial gradients of all three velocity components $dV_R/dR$, $dV_\theta/dR$ and $dV_Z/dR$. The values of these parameters are in good agreement with those obtained in  \citep{Siebert2011, Nelson2022} and also in  works \citep{Fedorov2021, Fedorov2023}, where these parameters were obtained in the framework of a linear model for the velocity field decomposition. 

The radial gradient of the Galactic expansion velocity $dV_R/dR$ is interpreted as the {\it contraction} ({\it expansion}) of the stellar system along the Galaxy radius. 
It may be the result of dynamical perturbations generated by the substructures: Galactic bar, spiral arms and warp. The value of this kinematic parameter in the Galactic plane varies within $\pm$10 \kmskpc. The detailed analysis of the kinematic parameter $dV_R/dR$ provides  a way to determine the parameters of the spiral structure, as recently shown by \citet{Denyshchenko2024}.
 
The kinematic parameter $dV_{\theta}/dR$ is a radial gradient of the circular velocity component, which shows the value of the {\it slope} of the Galaxy rotation curve at a given point. As can be seen from Fig.~\ref{fig:d_dR}, the value of this parameter varies between $-30$\,\kmskpc near the Galaxy center to 10 \kmskpc at the Galactocentric distances of 10 kpc and more. Note that $V_\theta$ has the opposite sign with respect to the Galactic rotation $V_\text{rot}$ (see Eqs. (\ref{eq:dVtheta_dR_om}) and (\ref{eq:Vrot_dVR})), hence the negative value of the derivative $dV_\theta/dR$ represents an increase in $V_\text{rot}$ and vice versa. Figure \ref{fig:d_dR} shows a rapid increase in the rotation velocity from the Galaxy center up to $R\simeq$ 5 kpc. A slow decrease in the Galaxy rotation velocity is seen from 7.5~kpc, corresponding to a positive value of $dV_\theta/dR$.
The slope $dV_\theta/dR$ appears in good agreement with the map of the azimuthal velocity $V_\theta$ (fig. \ref{fig:Velocity_residual}) and with the rotation velocity curve (fig. \ref{fig:Vrot}).

\begin{figure*}
\includegraphics[width = 58mm]{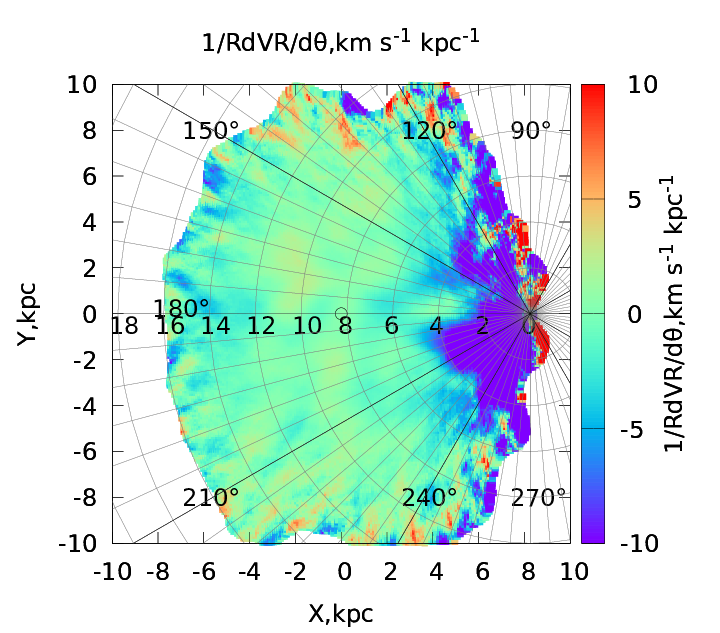}
\includegraphics[width = 58mm]{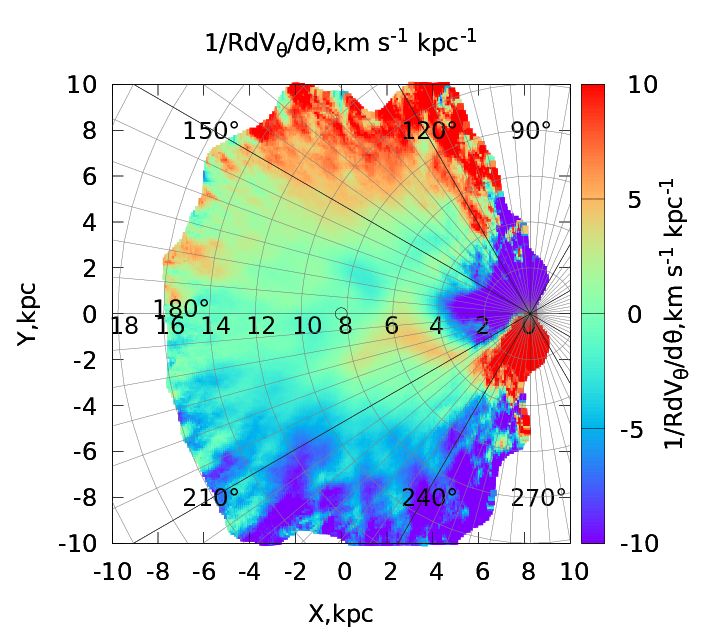}
\includegraphics[width = 58mm]{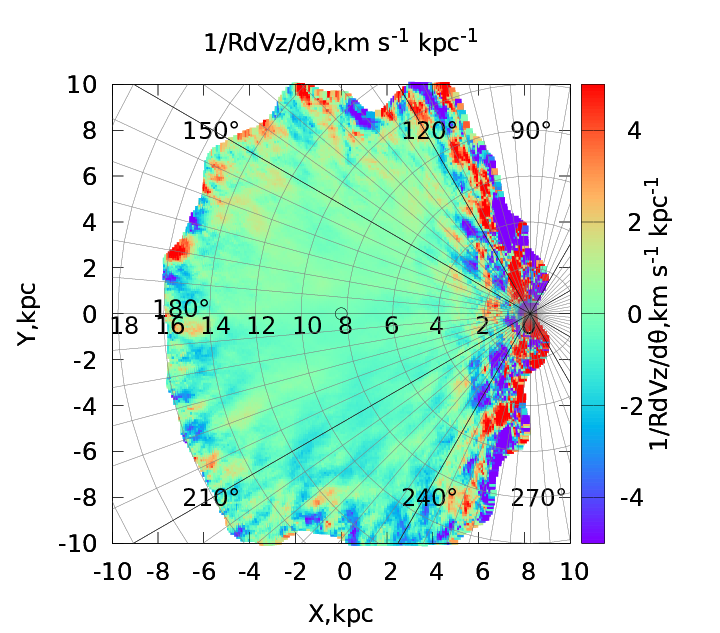}
\caption{The azimuth gradient of the velocity component (from left to right, respectively, $\frac{1}{R}\frac{dV_R}{d\theta}, \frac{1}{R}\frac{dV_{\theta}}{d\theta}, \frac{1}{R}\frac{dV_Z}{d\theta})$ as a function of the Galactic coordinates.}
\label{fig:d_dtheta}
\end{figure*}

In the right panel of figure \ref{fig:d_dR} the value of the radial gradient of the vertical velocity $dV_Z/dR$ is close to zero in the Solar neighbourhood and less than 2 \kmskpc\, in other regions. This means that the first order derivatives of the vertical velocity component along the Galactic radius and azimuth directions are very small values, as we can see in the right panels of Figs \ref{fig:d_dR} and \ref{fig:d_dtheta}.
This corresponds to a smooth and slow increase of the vertical velocity with increasing Galactocentric distance. Only at a Galactocentric distance greater than 12 kpc the value of $dV_Z/dR$ is equal to +3 \kmskpc, which is a signature of the Galactic warp.

\subsection{The azimuthal gradient of the stellar velocity field}

Using our method, for the first time we have obtained the estimates of the azimuth gradients of the velocity components.
In the assumption of an axisymmetric potential, all velocity gradients along the changing angle $\theta$ must be zero. 
However, as can be seen from Figure \ref{fig:d_dtheta}, the dependencies of $dV_R/d{\theta}$ on the left and $dV_{\theta}/d{\theta}$ in the middle panel are non-zero. 

The kinematic parameter $(1/R)dV_R/d{\theta}$ is the azimuthal gradient of the radial velocity component. As can be seen from Fig.~\ref{fig:d_dtheta} ({\it left}), this gradient is a function of coordinates and it reaches values of \mbox{$\pm$(3-5) \kmskpc}. The estimation of this parameter is very important to determine the Galaxy rotation velocity, as well as to find the correct value for the Oort constants $A$, $B$, $C$ and $K$, as we can see from the equations (\ref{eq:Oort_A}), (\ref{eq:Oort_B}), (\ref{eq:Oort_C}) and (\ref{eq:Oort_K}).

Since the parameter $(1/R)dV_R/d{\theta}$ cannot be directly estimated by the O--M model, it is conveniently put to zero in the frame of the axisymmetric rotation model.
In Fig. \ref{fig:Vrot} we can see the difference between the Galaxy rotation curves obtained in the framework of the O--M model ({\it blue dots} - equation (\ref{eq:Vrot})), and with the model that takes into account $(1/R)dV_R/d{\theta}$  ({\it green dots}) as shown in equation (\ref{eq:Vrot_dVR}). This difference in the Galactic rotation velocity reaches $\pm$10 \kms in the range of 6-12 kpc in Galactocentric distances. At other distances, if we do not consider the parameter $(1/R)dV_R/d{\theta}$, the Galaxy rotation velocity component has a significant distortion reaching several tens of \kms. We can conclude that not taking into account the parameters $(1/R)dV_R/d{\theta}$ introduces a significant distortion in the determination of the Galaxy rotation curve.

The middle panel of Fig. \ref{fig:d_dtheta} shows the azimuth gradient of the circular velocity component $(1/R)dV_{\theta}/d{\theta}$. As mentioned above, this parameter cannot be determined from the O--M model without additional assumptions. On the contrary, the approach we have implemented allows us to determine $(1/R)dV_{\theta}/d{\theta}$ from equation (\ref{eq:derivativesVth}) without additional conditions. 

In figure \ref{fig:d_dtheta}, we observe a smooth change of the kinematic parameter $(1/R)dV_{\theta}/d{\theta}$ as a function of the $\theta$ angle. Thus, at distances greater than 8 kpc from the Galactic center and for \mbox{$\theta\simeq 210^{\circ}$}, the values of this parameter are about $-5$ \kmskpc and increase in the direction of the Galactic rotation up to a value of +5 \kmskpc for $\theta\simeq 150^{\circ}$. Along the line between the Sun and Galactic anticenter, the value of $(1/R)dV_{\theta}/d{\theta}$ is close to zero.
An asymmetric azimuthal gradient of the circular velocity with opposite slope for the line with $\theta > 215^{\circ}$ and $\theta < 215^{\circ}$ is well seen in the inner disc ($R\la 4$ kpc). This is an additional kinematic signature of the Galactic bar already evidenced in Fig.~\ref{fig:Velocity_residual}.

We also notice that the middle panel of figure \ref{fig:d_dtheta} shows an opposite velocity gradient $(1/R)dV_{\theta}/d{\theta}$, both a positive and a negative for   $Y>+7$ kpc and $Y<-7$ kpc, respectively. This result  derives from the higher rotation velocity, $V_{\rm rot}\equiv -V_\theta \ga 250$ \kms, estimated in the outer regions ($|Y|> 7$ kpc) with respect to the mean velocity  $V_{\rm rot}\sim 220$ \kms in the Solar neighbourhood (see fig. \ref{fig:Velocity_residual}). The high value of the rotation velocity in the outer regions is likely overestimated because of the  incompleteness of our stellar catalog close to the Galactic plane. This means that the azimuthal velocity  $(V_\theta)_{xy}$ at $Z=0$ may result biased because the best fit of the second order Taylor expansion (\ref{eq:derivativesVth} is mainly constrained by stars outside the Galactic plane.

As described earlier, the azimuthal gradient of the vertical velocity $(1/R) dV_Z/d\theta$ is practically zero in the entire plane of the Galaxy. This parameter is associated with the kinematic evidence of the Galactic warp. \cite{Miyamoto1998} and \cite{Zhu2000} determined the following value $(1/R) dV_Z/d\theta$ = $3.8\pm$1.1 \kmskpc based on the proper motions of O-B stars in the HIPPARCOS catalogue. At the same time, \cite{Mignard2000} also found no confirmation of the Galaxy warp based on the HIPPARCOS data. \cite{Vityazev2012}, applying the Vector Spherical Harmonics (VSH) and Zonal Vector Spherical Harmonics (ZVSH) methods based on the stellar proper motions of the Tycho2 and UCAC3 \citep{Zacharias2010} catalogues, obtained statistically insignificant parameter values $(1/R)dV_Z/d\theta$ both for the whole sphere and separately for the Northern and Southern hemispheres. From our analysis, the vertical velocity does not show dependence on the Galaxy coordinates $R$ and $\theta$ as we can see in Fig.s \ref{fig:d_dR}, \ref{fig:d_dtheta} and \ref{fig:d2_dtheta} (right panels).  

\begin{figure*}
\includegraphics[width = 58mm]{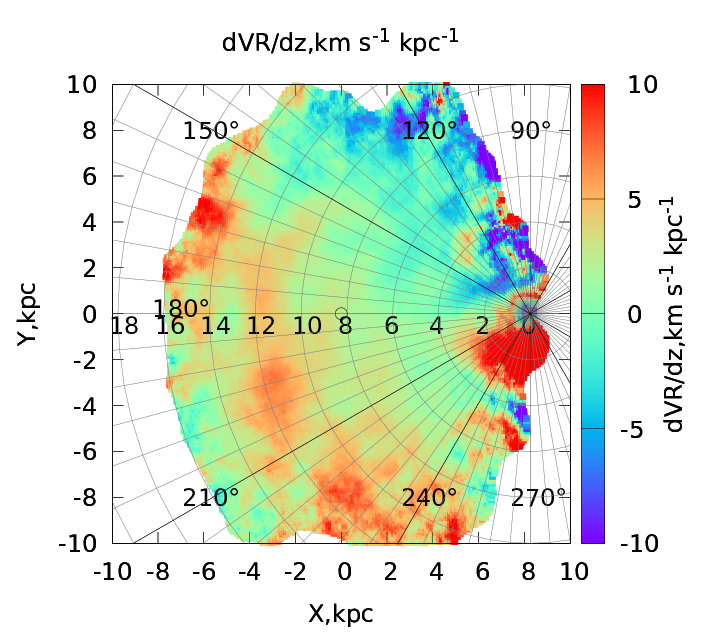}
\includegraphics[width = 58mm]{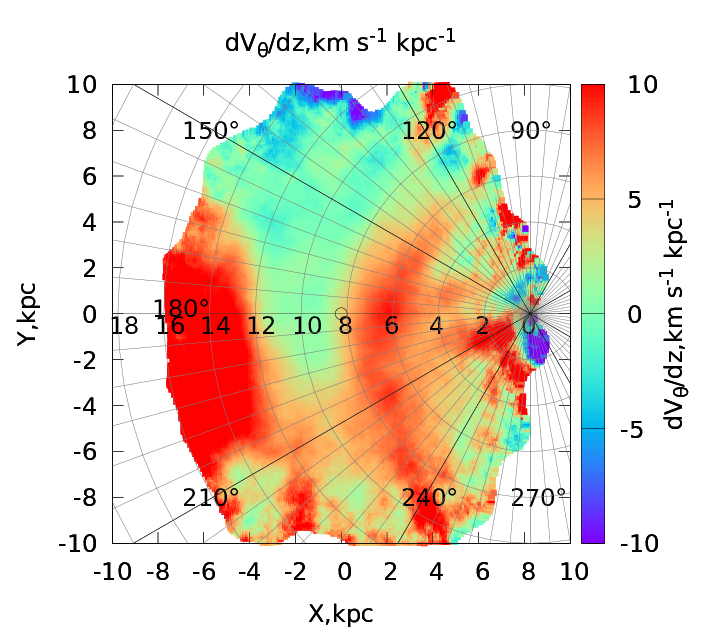}
\includegraphics[width = 58mm]{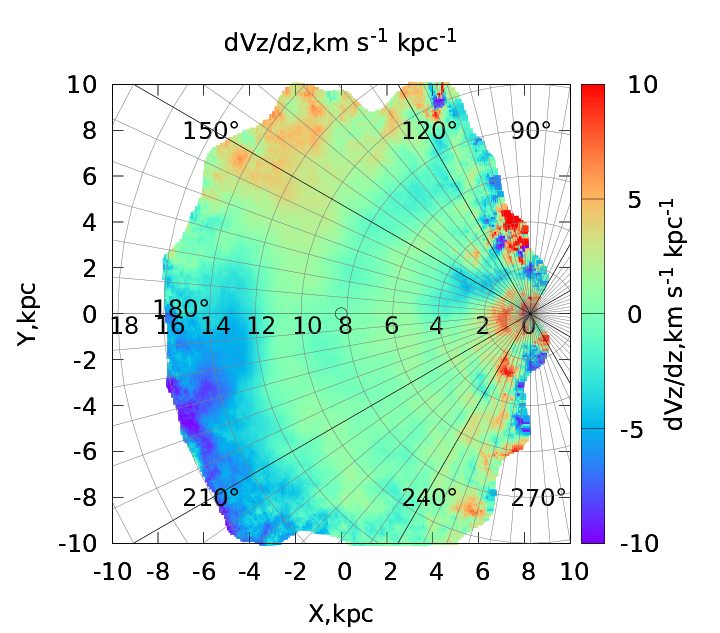}
\caption{The vertical gradient of the velocity component (from left to right, respectively, $\frac{dV_R}{dZ}, \frac{dV_{\theta}}{dZ}, \frac{dV_Z}{dZ})$ as a function of the Galactic coordinates. The figures show the value of the difference between the vertical gradient of the stellar velocity field in the Northern and Southern hemispheres of the Galaxy.}
\label{fig:d_dz}
\end{figure*}

\subsection{The vertical gradient of the stellar velocity field}

In this section we consider the first order vertical gradients of the velocity components $dV_R/dZ$, $dV_\theta/dZ$, and $dV_Z/dZ$ illustrated in Fig.~\ref{fig:d_dz}. \cite{Vityazev2012} showed that the kinematic parameters $dV_R/dZ$ and $dV_\theta/dZ$, determined separately for the Northern and Southern hemispheres using Zonal Vector Spherical Harmonics (ZVSH), have different signs but they are close in absolute values. However, if we consider the whole sphere, the values of these parameters are close to zero. In this work, we use a symmetric distribution of stars in both the Northern and Southern Galactic hemispheres. These parameters describe the difference between the vertical gradients in the Northern and Southern hemispheres -  antisymmetric component of the vertical gradient. We estimate the values of the symmetric vertical gradient component using second order derivatives with respect to $Z$ (see details in the next section).

The values of the kinematic parameter $dV_R/dZ$ vary within the range $\pm$10 \kmskpc and show a clear dependence on the coordinates $R$ and $\theta$ (Fig. \ref{fig:d_dz}, {\it left}). The values of $dV_R/dZ$ are positive and show a slight increase with Galactocentric distance in the region with $\theta > 160^\circ$. This means that for this region of the Galaxy the absolute value of the vertical gradient of the Galactic expansion is larger in the Northern hemisphere than in the Southern. This conclusion can also be made from the results of \cite{Vityazev2012}, \cite{Nelson2022} and \cite{Wang2023}. For example, \cite{Vityazev2012} used the data of UCAC3 catalogue from \citep{Zacharias2010}. They obtained the values of $dV_R/dZ$ in the region of the Sun \mbox{$+22.2 \pm0.5$ \kmskpc} and $-10.8\pm0.5$\kmskpc for the Northern and Southern hemispheres, respectively. In the Galaxy region with $\theta<160^\circ$, the kinematic parameter $dV_R/dZ$ has a negative value and also shows an increase in absolute value with an increase in $R$.

Fig. \ref{fig:d_dz} ({\it middle}) shows the non-symmetry with respect to the Galactic plane of the vertical gradients of the Galaxy rotational velocity $dV_\theta/dZ$ in the Northern and Southern hemispheres. 
The first proof of slowing down of the Galaxy rotation velocity with increasing distance from the Galactic middle plane was given by \cite{Majewski1993, Girard2006}, \cite{Makarov2007} with estimation of the corresponding parameter as +20 \kmskpc. 
\cite{Vityazev2012}, using data from the UCAC3 catalogue, determined with high accuracy the values of the vertical velocity gradient of Galactic rotation of $dV_\theta/dZ = +26.9 \pm0.6$ \kmskpc and $-52.1\pm0.6$ \kmskpc in the Northern and Southern hemispheres, respectively. It was made separately via the Zonal-VSH method. 
Later, \cite{Velichko2020} found similar results using the stellar proper motions of the catalogues: \emph{Gaia} DR2 \citep{Katz2018} and PMA \citep{Akhmetov2017}.
The authors of these papers did not interpret the differences in the absolute values of $dV_{\theta}/dZ$ for each hemisphere; they averaged this value and found its lower and upper estimations. Fig. \ref{fig:d_dz} ({\it middle}) shows the difference in the absolute values of the vertical gradient of the Galaxy rotation velocity in the Northern and Southern hemispheres. In the region close to the Sun, the value of this parameter is about +7 \kmskpc, which means that the slowdown velocity of the Galaxy rotation in the Northern hemisphere is less than in the Southern one. We can make such a conclusion from the analysis of \citep{Nelson2022}, their figures 4 and 6. We can see the same behavior of the asymmetry of the vertical gradient at Galactocentric distances from 0 to 8.5 kpc and beyond 12 kpc for almost all $\theta$ angles. Possibly, the asymmetry of the vertical gradient of the Galaxy rotation in the region with large Galactic distances beyond 12 kpc can  be also explained by the presence of the Galactic warp. 

Fig. \ref{fig:d_dz} ({\it right}) shows the dependence of the difference (non-symmetry with respect to the Galactic plane) of the vertical gradient of the vertical stellar velocity between the Northern and Southern hemispheres. As can be clearly seen in this figure, this kinematic parameter is close to zero near the Sun. The value of $dV_Z/dZ$ begin to deviate significantly from zero only at distances greater than 10~kpc and its behavior looks like the parameter $dV_{\theta}/dZ$ but with an opposite sign. We can also see an increase in the absolute value of this parameter with $R$ and a sign change at $\theta=160^\circ$.

If we do not consider the local variations of the parameter $dV_\theta/dZ$ in Fig. \ref{fig:d_dz} ({\it middle}), this parameter also shows a significant change near the line with  $\theta = 160^\circ$. Thus, on this line we see a sign change in all kinematic parameters $dV_R/dZ$, $dV_{\theta}/dZ$, and $dV_Z/dZ$, showing an asymmetry of the vertical gradients of the velocity components relative to the Galaxy middle plane (differences in the Northern and Southern hemispheres). Possibly, these vertical perturbations of the stellar velocity field are due to the influence of the Sagittarius galaxy and/or the Large Magellanic Cloud \citep{Garavito-Camargo2019} and \citep{Vasiliev2021}.

\begin{figure}
\includegraphics[width = 88mm]{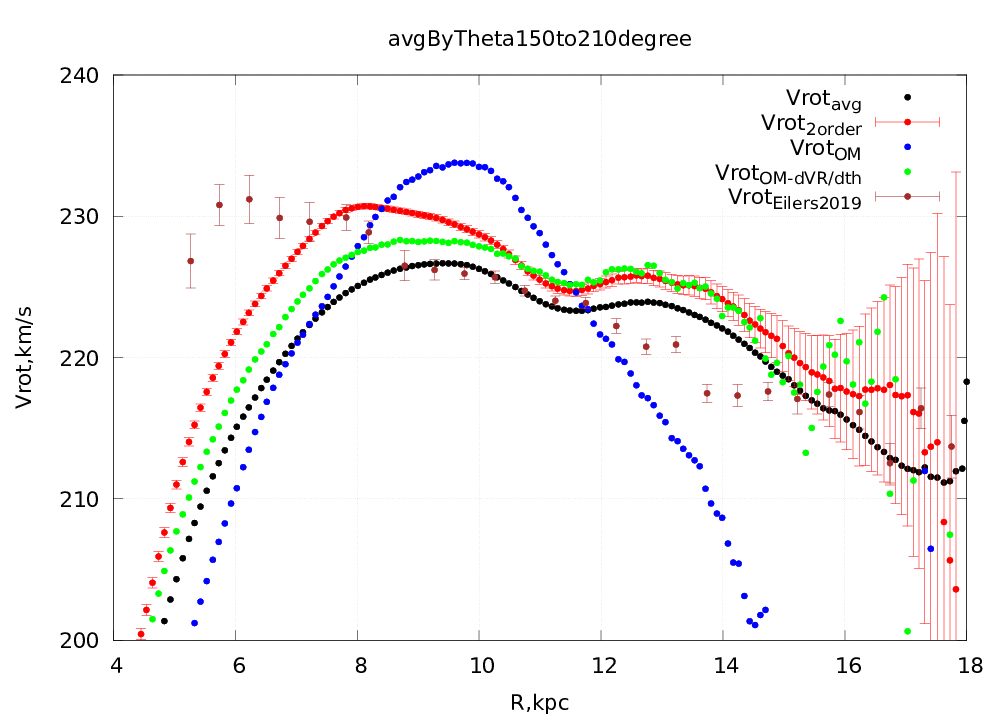}
\caption{ {\it Black dots} correspond to the value of the Galaxy rotation curve obtained by averaging the stellar velocity component $V_\theta$ for all "fictitious stars" without using any kinematics model. The Galaxy rotation curves calculated using our method from equation (\ref{eq:derivativesVth}) ({\it red dots}), using O--M model with the Oort parameters $A$ and $B$ of the formula (\ref{eq:Vrot}) ({\it blue dots}). We also present the rotation curve taking into account $\partial V_R/\partial \theta$ using eq.~(\ref{eq:Vrot_dVR}) ({\it green dots}).
}
The values of the Galaxy rotation curves are obtained by averaging the corresponding data in the range of angles $\theta$ from $150^\circ$ to $210^\circ$ ($\pm 30^{\circ}$ about the direction of the Galactic Centre -- the Sun -- the Galactic anticentre).We compare the rotation curves derived in this work with {\it circular} velocity curve $V_c(R)$ derived by \citet{Eilers2019} ({\it brown dots}).
\label{fig:Vrot}
\end{figure}

\subsection{Azimuthal stellar velocity as a function of radii: the Galactic rotation curve}

Figure \ref{fig:Vrot} shows the Galaxy rotation curves obtained by different methods from the azimuthally averaged values of the rotation velocity component in the range $\theta$ from $150^\circ$ to $210^\circ$. The {\it black dots}  correspond to the value of the Galaxy rotation curve obtained by averaging the stellar velocity component $V_\theta$ for all "fictitious stars" without using any kinematic model. The ({\it blue points}) represent the Galaxy rotation curve obtained from the Oort parameters $A$ and $B$ of equation (\ref{eq:Vrot}), by assuming an axisymmetric rotation and $dV_R/d\theta$ = 0. The ({\it green dots}) show the  Galactic rotation curve  from equation  (\ref{eq:Vrot_dVR}), which takes into account the non-axisymmetric rotation $dV_R/d\theta$ from equation (\ref{eq:derivativesVR}).
Figure \ref{fig:Vrot} also shows the Galaxy rotation curve $V_{rot}=-V_{\theta} $ derived from equation (\ref{eq:derivativesVth}) ({\it red dots}). We note that these values are systematically larger in the range of Galactocentric distances from 4 to 9 kpc than those calculated by other methods. This is due to the vertical velocity gradient that is modeled by the quadratic term $\partial^2V_\theta/\partial Z^2$, while the other methods can not correct the systematics due to the vertical gradient on the circular velocity, as the first order expansion of the circular stellar velocity along $Z$ is insufficient. The values of the Galactic rotation velocity and their errors are given in Tables \ref{tab:galaxy_curve_1} and \ref{tab:galaxy_curve_2}.  

Thus, we argue that our model described by equation (\ref{eq:derivativesVth}) ({\it red dots}) is  able to determine the mean Galactic rotation velocity in the Galactic plane ($Z=0$) most accurately, because it takes into account both the influence of the vertical gradient and the non-axisymmetric rotation.

\begin{figure*}
\includegraphics[width = 58mm]{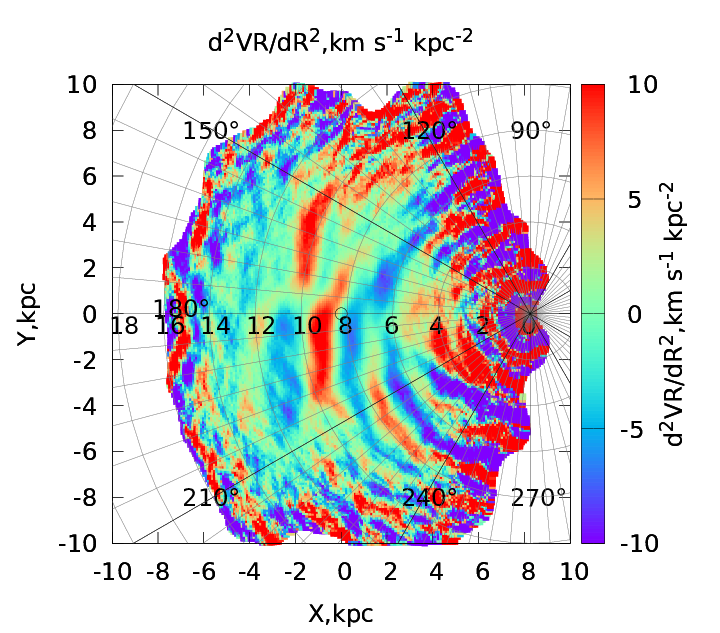}
\includegraphics[width = 58mm]{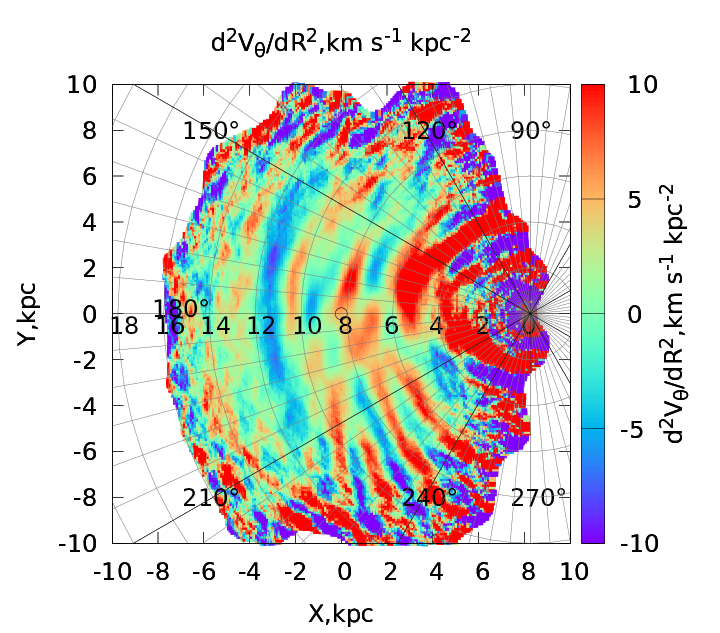}
\includegraphics[width = 58mm]{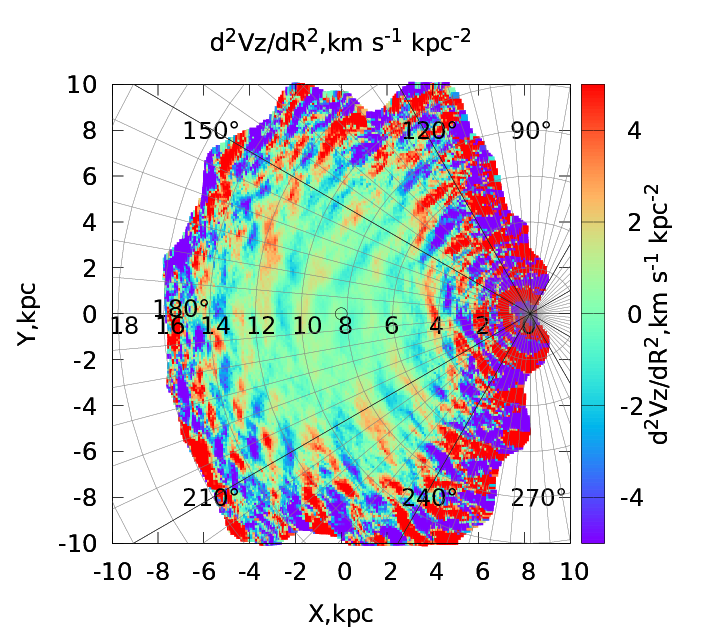}
\caption{The second order derivatives of the stellar velocity field in the radial direction. The wavelike behavior of the stellar velocity field along the radial direction.}
\label{fig:d2_dR}
\end{figure*}

\begin{figure*}
\includegraphics[width = 58mm]{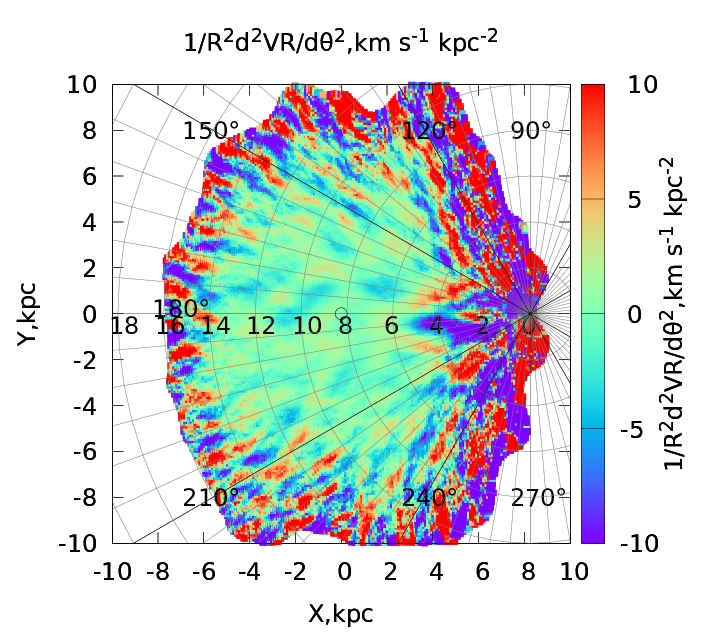}
\includegraphics[width = 58mm]{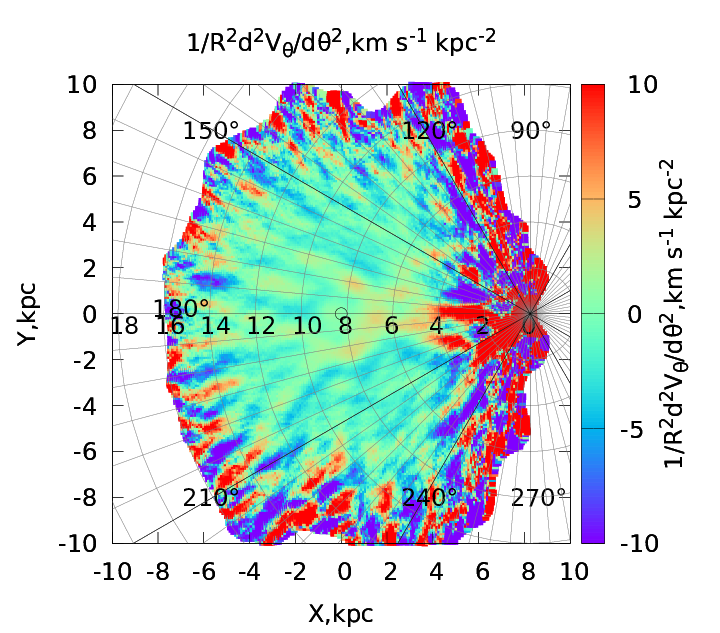}
\includegraphics[width = 58mm]{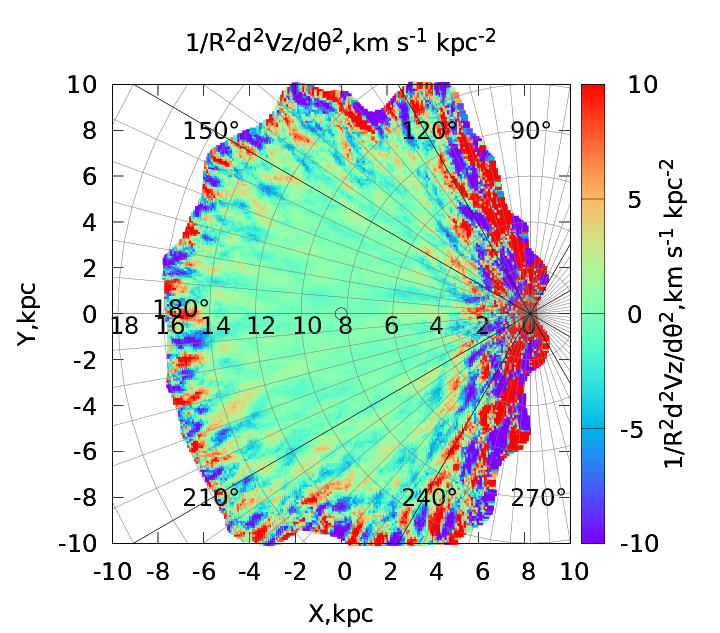}
\caption{The second order derivatives of the stellar velocity field along of the azimuthal angle. The wavelike behavior of the stellar velocity field along the azimuth direction.}
\label{fig:d2_dtheta}
\end{figure*}

\section{Second order directional derivative of the stellar velocity field}
\label{sec:Second_order} 

In this paper, we present for the first time the results of the expansion of the stellar velocity field into the Taylor series up to the second order for \emph{Gaia} data. 
As we show in the Appendix \ref{sec:app1} of Fig. \ref{fig:edV2} and \ref{fig:edV2mix}, the accuracy of the second derivatives is somewhat lower than that of the first, but all these parameters are estimated quite reliably at the level of $\pm$0.5 -- 4 \kmskpcc at the heliocentric distance up to 5-6 kpc. We give below a short description and physical interpretation of only some of the most significant second derivatives.

\subsection{Second order derivatives of the stellar velocity field in the radial direction }

In our opinion, the most interesting are the second derivatives of the stellar velocity field along the Galactic radius $d^2V_R/dR^2$, $d^2V_{\theta}/dR^2$, and $d^2V_Z/dR^2$ shown in figure \ref{fig:d2_dR}.

As it can be seen in the left and middle panels of figure \ref{fig:d2_dR}, these second order derivatives for radial $V_R$ and azimuthal $V_{\theta}$ components of the stellar velocity field show significant changes within $\pm$10 \kmskpcc. Some patterns in the form of alternating ring-shaped structures are also clearly visible in the figure \ref{fig:d2_dR}. We see alternating blue and red annular structures reflecting the local maxima and minima of the Galactic expansion velocity $V_R$ and rotation velocity $V_\theta$ on the scale 1-2 kpc, corresponding to the spatial resolution of our velocity maps. In our case, this behaviour is an analog of the Nyquist frequency. 

Similar ring-like patterns are also present in the vertical velocity component $d^2V_Z/dR^2$ (Fig. \ref{fig:d2_dR}, right panel). However, their amplitudes are significantly smaller than those observed for the other two velocity components, reaching values no higher than $\pm$3 \kmskpcc.

The second order derivatives of the stellar velocity field in the radial direction show a wave-like dependence on Galactic distance $R$. 
All these ring-like signatures evidenced in the  second order derivatives of the stellar velocity field in the radial direction are probably related to kinematic substructures, such as the spiral arms and bulge, and/or by the propagation of density, bending and breathing waves.

\subsection{Second order derivatives of the stellar velocity field along the azimuthal angle}

Figure \ref{fig:d2_dtheta} shows the second order derivatives along the azimuthal angle. As can be seen from the left and middle panels of Fig.\ref{fig:d2_dtheta}, the parameters $(1/R^2)d^2V_R/d\theta^2$ and $(1/R^2)d^2V_{\theta}/d\theta^2$ show a weak change that do not exceed $\pm$3 \kmskpcc and do not have a clear dependence on Galactic coordinates. The parameter $(1/R^2)d^2V_Z/d\theta^2$ has an insignificant value practically in the whole range of Galactic coordinates (the right panel of Fig. \ref{fig:d2_dtheta}). 

Thus, in the middle plane of the Galaxy, no clearly discernible wavelike behavior of the stellar velocity field is observed. Insignificant changes of the parameters $(1/R^2)d^2V_R/d\theta^2$ and $(1/R^2)d^2V_{\theta}/d\theta^2$ characterize the local change of stellar velocities and do not describe the global processes in the Galaxy.

\subsection{Second order derivatives of the stellar velocity field in the vertical direction}

\begin{figure*}
\includegraphics[width = 58mm]{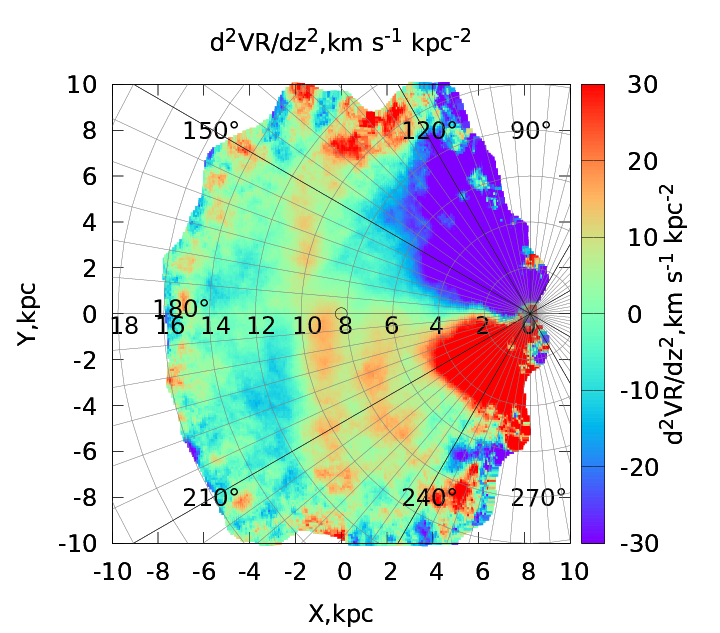}
\includegraphics[width = 58mm]{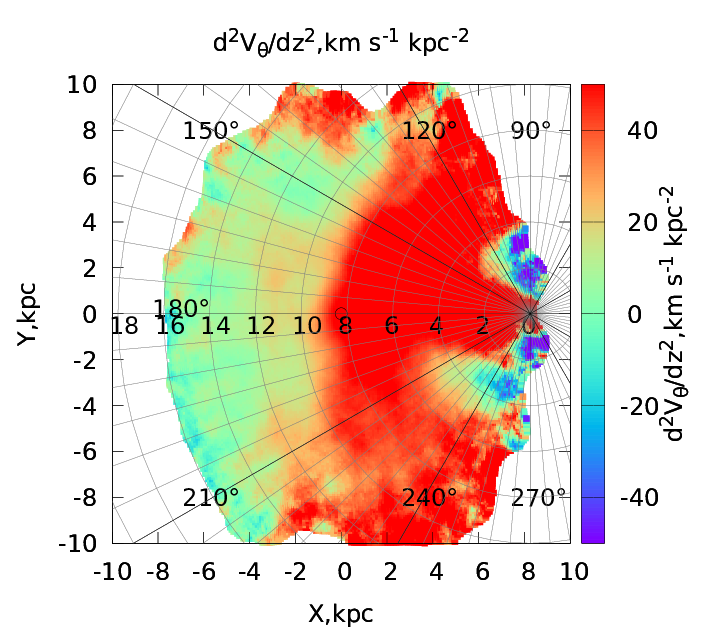}
\includegraphics[width = 58mm]{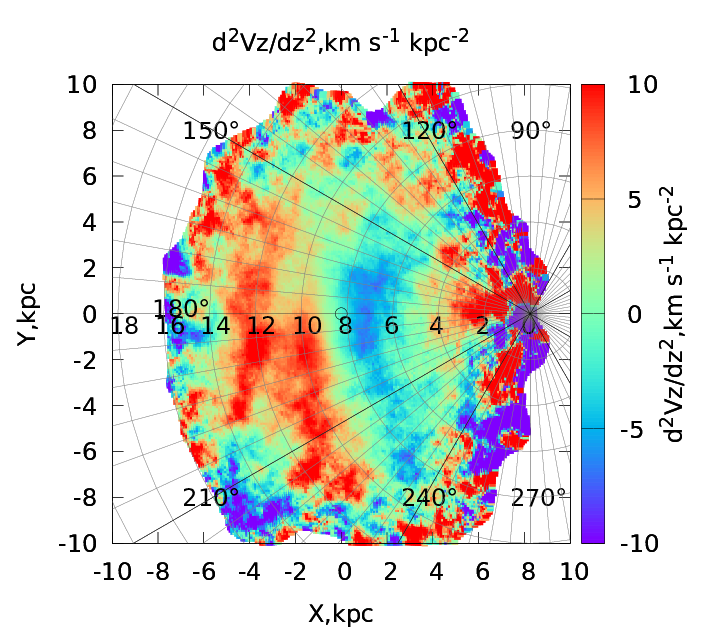}
\caption{The second order derivatives of the stellar velocity field in the vertical direction. The symmetry part of the vertical gradient  of the components $V_R, V_{\theta}, V_Z$ stellar velocity field with respect to the Galactic middle plane.}
\label{fig:d2_dz}
\end{figure*}

The main value of the vertical gradient of stellar velocities results from the symmetry in absolute value and the difference in sign of the vertical gradient of stellar velocity fields in the Northern and Southern hemispheres. As \cite{Vityazev2012} have shown, using a symmetric sample of stars relative to the Galactic middle plane, it is impossible to estimate the value of the vertical gradient  within the framework of the O--M linear model or the VSH method. 
Our models account for the possibility that not only the sign of the vertical gradient (symmetric component) but also its absolute value can change from the Northern to the Southern hemisphere -- the antisymmetric component of the vertical gradient. 
Fig. \ref{fig:d2_dz} shows the values of the parameters $d^2V_R/dZ^2$, $d^2V_{\theta}/dZ^2$ and $d^2V_Z/dZ^2$ corresponding to the symmetric component of the vertical gradient of the stellar velocity fields.
As mentioned above, we have already considered the antisymmetric component $dV_R/dZ$ and $dV_{\theta}/dZ$ in Fig. \ref{fig:d_dz}.

The middle panel of Fig. \ref{fig:d2_dz} shows the map of the second order derivatives of the Galactic rotation velocity in the vertical direction. The value of $d^2V_{\theta}/dZ^2$ reaches 50 {\kmskpcc} in the Solar neighbourhood, that corresponds to an asymmetric drift of about $-25$ \kms at $|Z|= 1$ kpc. Such value is pretty consistent with the vertical gradient $dV_\mathrm{rot}=-20 \pm 6$ \kmskpc and $-22$ \kmskpc derived respectively by \citet{Makarov2007} and \citet{Levine2008}, who analysed stellar samples close to the Galactic plane ($|Z|<1$ kpc), as in our case.

In the range of Galactocentric distances $R$ between 4 and 9 kpc, the value of the second order derivatives of the Galactic rotation velocity $d^2V_{\theta}/dZ^2$ is practically independent of the $\theta$ angle and reaches a maximum value of about 50 \kmskpcc.
At the Galactocentric distances $R$ between 9 and 10 kpc, the value of the $d^2V_{\theta}/dZ^2$ quickly drops to 10 -- 20 \kmskpcc for all values of the $\theta$ angle. Also, in the outer region with $R > 10$ kpc, we see a smooth decrease almost to zero at a distance of about 15 kpc. 
Such difference  is due to the superposition of the thin disk with the slowly rotating thick disk in the inner disc, combined with the flaring of the thin disc in the outer regions \citep{Mackereth2017, Beraldo2020}. These results are consistent with previous studies carried out by \citet[Figs.\ 4 and 6]{Nelson2022}, \citet[Figs. 4-5]{Wang2023}, and \citet[Fig. E.1]{Recio-Blanco2023}.

The left panel of Figure \ref{fig:d2_dz} shows the Galactic map of the parameter $d^2V_R/dZ^2$ which, according to the physical meaning, is the second order derivative of the Galactic expansion velocity in the vertical direction. In the inner region of influence of the Galactic bar ($R$ < 5 kpc), we see the parameter values equal to +30 \kmskpcc for $\theta$ > 175$^\circ$, corresponding to an increase of the Galactic expansion velocity $V_R$ with increasing height $|Z|$ from the Galactic plane. In the region with an azimuth $\theta$<175$^\circ$, the value of the parameter $d^2V_R/dZ^2$ has the opposite sign $-30$\,\kmskpcc, indicating a decrease in the Galactic expansion velocity with increasing in $|Z|$. We also observe such behavior for $R$ in the range between 5 and 10 kpc, but the value is 2-3 times smaller than in the region of the Galactic bar.

We note that, for $R<5$ kpc, the effect of the parameter $d^2V_R/dZ^2$ is opposite to that of the Galactic expansion velocity $V_R$ shown in the left top panel of fig. \ref{fig:Velocity_residual}, where $V_R \ga +20$ \kms and $\la -20$\, {\kms} for $\theta$ < 185$^\circ$ and $> 185^\circ$, respectively.  This means that the radial stellar streams in the inner disc and bulge rapidly decrease with increasing $|Z|$. Thus, we point out that the second derivative  $d^2V_R/dZ^2$ can provide important constraints on the dynamical models of the Galactic bar.

In Fig. \ref{fig:d2_dz} (right panel) the parameter $d^2V_Z/dZ^2$ is the second order derivatives of the vertical stellar velocity in the vertical direction, which is positive in almost the whole Galactic plane and varies within the range from +3 to +10 \kmskpcc. Only in a small region with Galactic radii $R$ between 6 and 8 kpc and angles $\theta$ between 145$^\circ$ and 240$^\circ$, $d^2V_Z/dZ^2$ has negative sign and changes in the range from $-2$ to $-7$~\kmskpcc. We can conclude that, in most of the Galaxy, the vertical velocity $V_Z$ slightly increases with growing distance $Z$ from the Galactic plane. As we can see from Fig. \ref{fig:d2_dz}, the stars move on average from the Southern to the Northern hemisphere, slowing down slightly as they pass through the plane of the galaxy. The Sagittarius galaxy and/or the Large Magellanic Cloud may be responsible for these vertical motions. This will require a detailed study and dynamical modeling.

\subsection{The mixed second order derivatives of the stellar velocity field}

\begin{figure*}
{\includegraphics [width = 58mm] {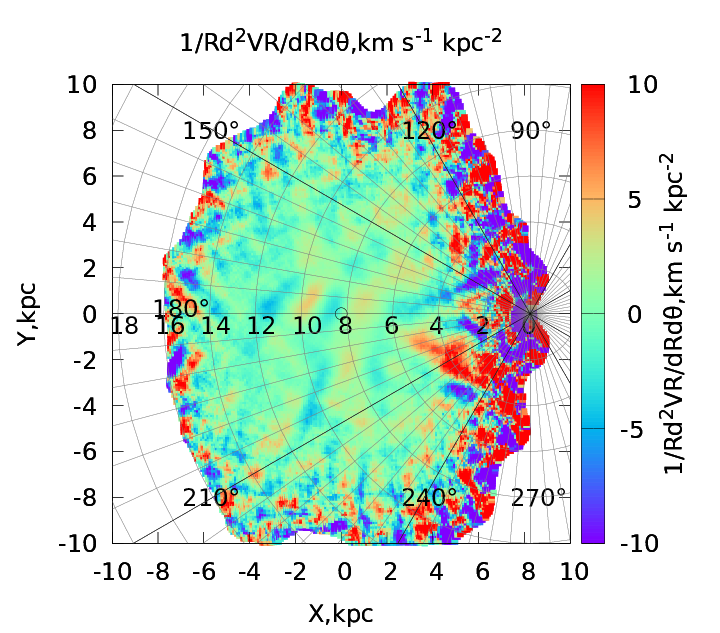}
 \includegraphics [width = 58mm] {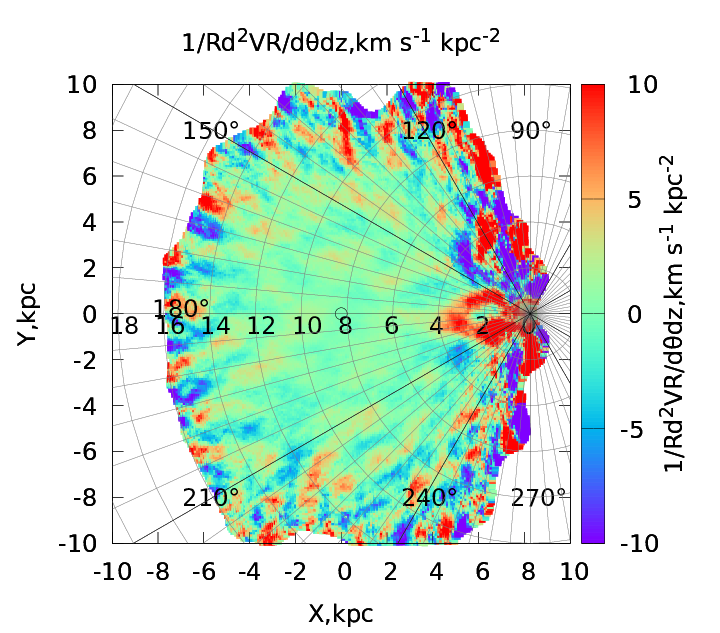}
 \includegraphics [width = 58mm] {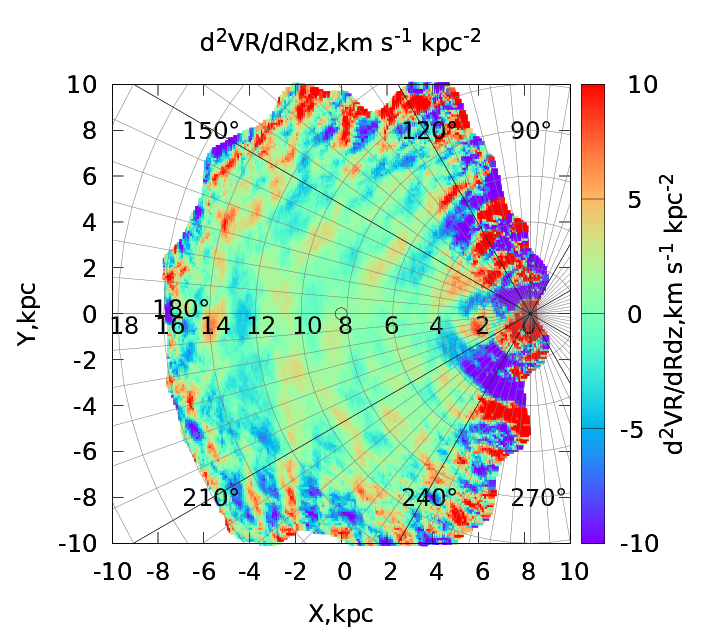}
 \includegraphics [width = 58mm] {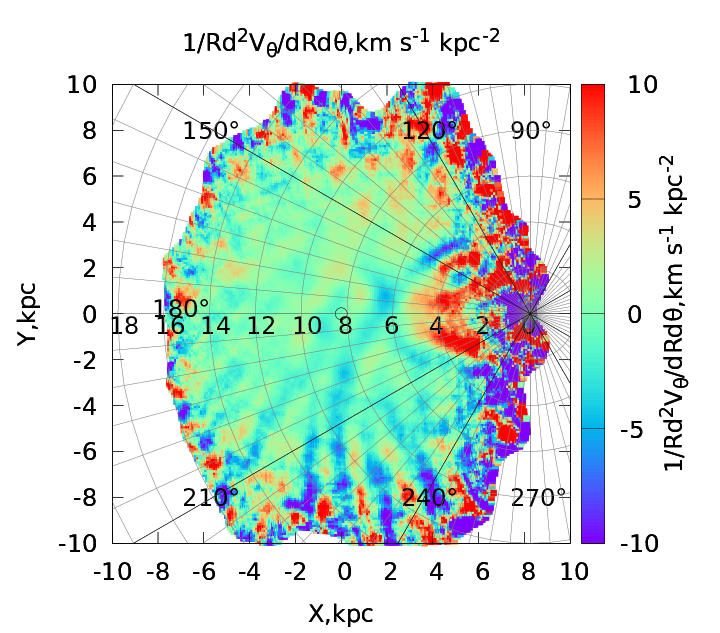}
 \includegraphics [width = 58mm] {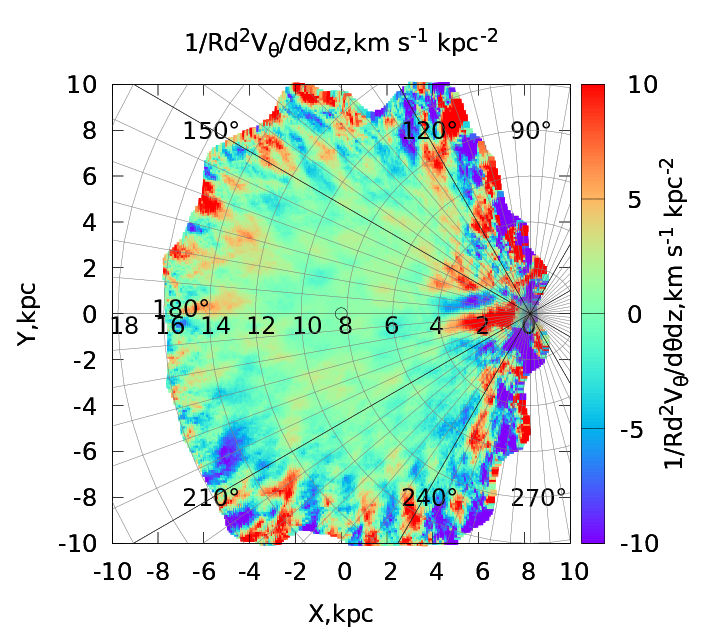}
 \includegraphics [width = 58mm] {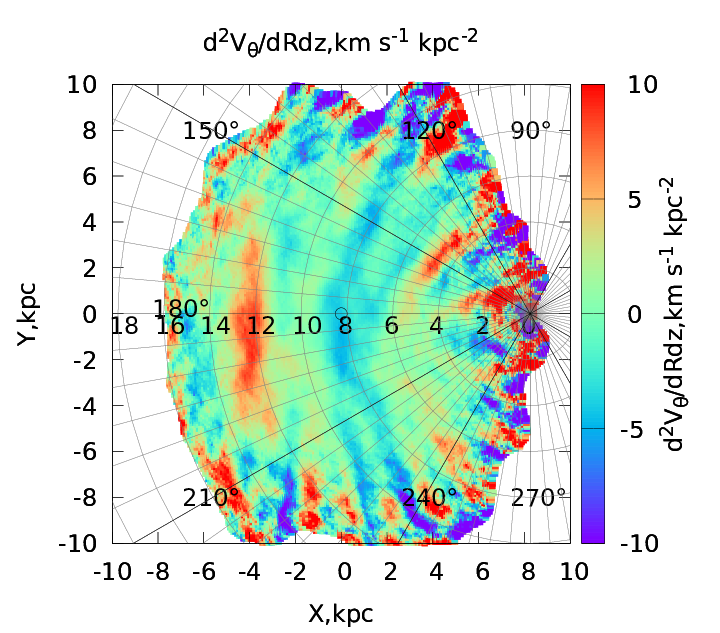}
 \includegraphics [width = 58mm] {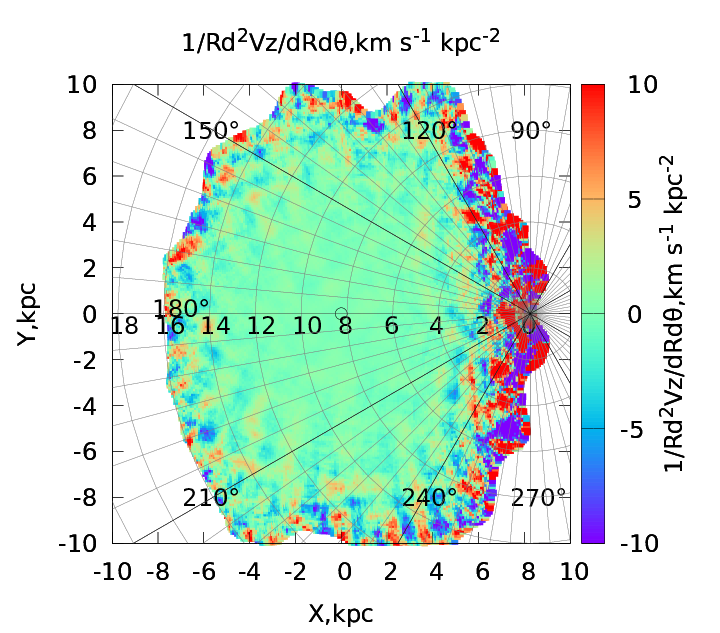}
 \includegraphics [width = 58mm] {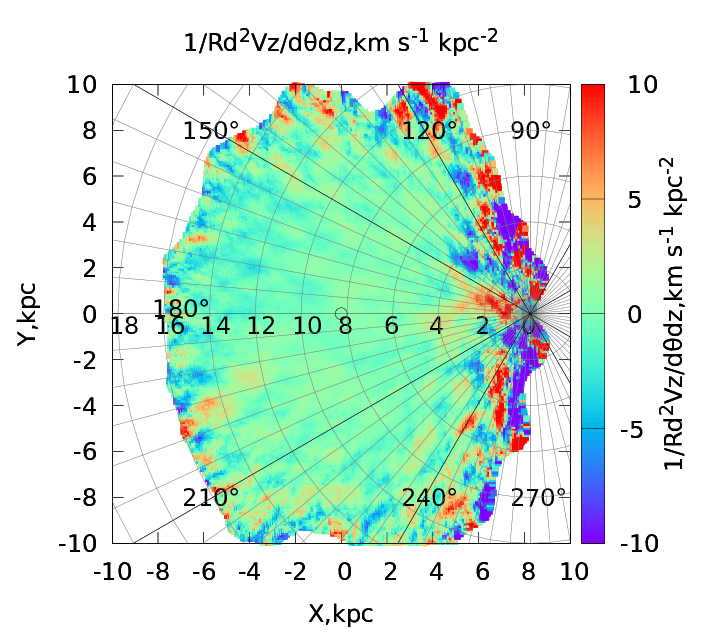}
 \includegraphics [width = 58mm] {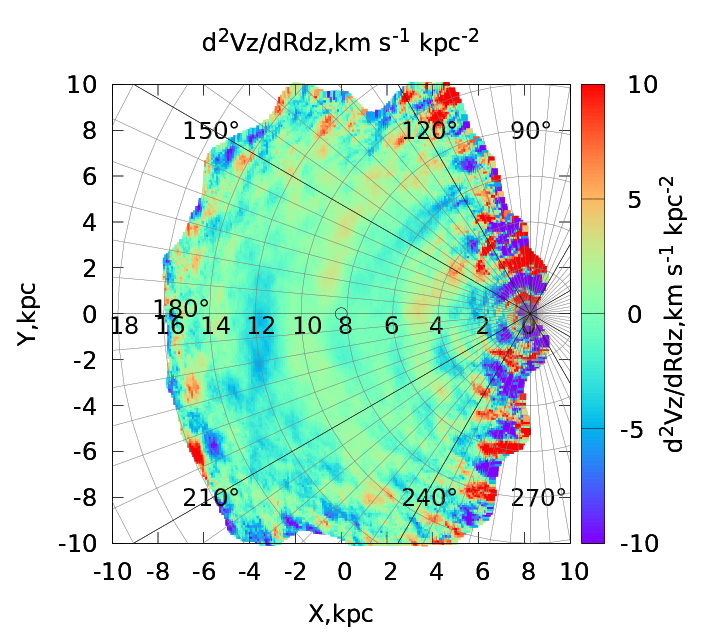}
 }
\caption{ The mixed second order derivatives of the stellar velocity field as a function of the Galactic coordinates.}
\label{fig:d2V_mix}
\end{figure*}

As we can see in figure \ref{fig:d2V_mix}, almost all mixed second order derivatives of the stellar velocity field have no obvious systematic behavior. The values of these parameters are close to zero within $\pm(3-4)$ \kmskpcc. Only two parameters $d^2V_{\theta}/dZdR$ and $d^2V_Z/dZdR$  at the Galactocentric distance of about 12 kpc and azimuth angles $\theta$ from $160^\circ$ to $200^\circ$ show a significant behavior and their values reach +10 \kmskpcc and $-5$\,\kmskpcc, respectively. We can interpret the parameter $d^2V_{\theta}/dZdR$ and $d^2V_Z/dZdR$ as gradients of the parameters $dV_{\theta}/dZ$ and $dV_Z/dZ$ along the Galactic radius $R$. As can be seen from Fig. \ref{fig:d_dz}, only parameters $dV_{\theta}/dZ$ and $dV_Z/dZ$ show a fast change in the region of Galactocentric distances $R$=12 kpc. Apparently, this is due to the influence of the Galaxy warp into the stellar velocity field, as shown in \citep{Poggio2018, Poggio2020}.

\section {Summary and conclusions}

In this paper, we have performed a kinematic analysis of the stellar velocity field for a sample of high luminosity stars from the \emph{Gaia} DR3 catalogue 
covering a large area of the Galactic plane, up to 10 kpc from the Sun. However, we have restricted our interpretation to a heliocentric distance of up to 6 kpc, where our results appear to be most reliable.

We used a new statistical method based on the decomposition of the stellar velocity fields into a Taylor series up to the second order in the Galactocentric cylindrical coordinate system. We have shown that the kinematic parameters obtained by this method are in good agreement with the results obtained by means of the Ogorodnikov-Milne model, Oort–Lindblad, and Gaussian process models \citep{Nelson2022}. 

Our method also provides additional kinematic parameters, including the gradients of the radial $\frac{1}{R}\frac{\partial V_R} {\partial \theta}$ and the azimuthal $\frac{1}{R}\frac{\partial V_\theta} {\partial \theta}$ velocity along the angle $\theta$, that cannot be derived by  the Ogorodnikov-Milne and Oort-Lindblad models. 
These two parameters are assumed to be zero by the axisymmetric Galactic models, so that significant non-zero values are a signature of non-axisymmetric structures.

We have computed the Oort constants $A$, $B$, $C$, $K$, and, for the first time, the second order partial derivatives of the stellar velocity field with respect to the Galactic coordinates $(R, \theta, Z)$.

The values of the kinematic parameters and the Oort constants for the Solar neighbourhood are listed in tables \ref{tab:kinematic_param} and \ref{tab:Oort_constant}, respectively.

We have estimated the values of the vertical gradient of the Galactic rotation, expansion and vertical velocity using the symmetric distribution of the stars with respect to the Galactic plane. We have shown that the stellar velocity fields have different behavior in the Northern and Southern hemispheres and are in good agreement with the results of \cite{Vityazev2012} obtained using ZVSH. 

We have also determined by different methods the Galactic rotation curve for Galactocentric distances from 4 kpc to 18 kpc by averaging Galactic azimuths in the range  \mbox{-30$^\circ$ < $\theta$  < +30$^\circ$} around the direction Galactic Centre -- Sun -- Galactic anticentre (fig. \ref{fig:Vrot}).

In conclusion, the presented method  provides a detailed description of the spatial stellar velocity field of the Milky Way and is able to reveal features related to non-axisymmetric substructures (e.g. spiral arms,  Galactic bar,  warp), as well as bending and breathing waves produced by satellite galaxies or other dynamical perturbations. Therefore, a detailed analysis of the new kinematic parameters in combination with dynamical modeling will significantly improve our understanding of the structure and evolution of our Galaxy.

The derived kinematic parameters and their errors can be made available to interested readers on personal request by e-mail: \href{mailto:akhmetovvs@gmail.com} {akhmetovvs@gmail.com}. 

\section{Acknowledgements}
This work has made use of data from the European Space Agency (ESA) mission \emph{Gaia} (\url{https://www.cosmos.esa.int/gaia}), processed by the \emph{Gaia} Data Processing and Analysis Consortium (DPAC,\url{https://www.cosmos.esa.int/web/gaia/dpac/consortium}). Funding for the DPAC has been provided by national institutions, in particular the institutions participating in the \emph{Gaia} Multilateral Agreement.

V. Akhmetov and E. Bannikova are very grateful to all Italian colleagues from different Astronomical Observatories of the Italian National Institute for Astrophysics (INAF) for the help during the evacuation from Kharkiv and for the following support in Italy.

The work of V. Akhmetov and E. Bannikova was supported under the special program of the NRF of Ukraine ”Leading and Young Scientists Research Support” – ”Astrophysical Relativistic Galactic Objects (ARGO): life cycle of active nucleus”, No. 2020.02/0346.
The authors also thanks Peter Fedorov, Alexey Butkevich and Ronald Drimmel for reading this paper and suggesting improvements and for useful discussions and comments.

We sincerely thank the anonymous Reviewer for carefully reading the paper, for very useful comments, and most importantly, for constructive suggestions, as his/her comments led us to improve the work.

\section*{Data availability}
\addcontentsline{toc}{section}{Data availability}
The catalogue data used are available to the reader in a standardised format via the CDS (https://cds.u-strasbg.fr).
The software code used in this work and the estimation of the kinematic parameters can be made available on personal request by e-mail:
\href{mailto:akhmetovvs@gmail.com} {akhmetovvs@gmail.com}.

\appendix
\section{The distribution of random errors of the kinematic parameters}
\label{sec:app1}

In the estimation of the kinematic parameters, individual errors of the measured astrometric parameters were not taken into account. The standard errors of the model kinematic parameters were estimated using the diagonal elements of the inverse matrix of the normal equations and the unit weight error. We drew contour lines with a step size of 0.5 on the maps representing the random errors of the kinematic parameter estimates in the Galactic middle plane derived by solving the LSM  the equations (\ref{eq:derivativesVR}), (\ref{eq:derivativesVth}) and (\ref{eq:derivativesVz}). $R_{\odot}$=8.249~kpc and shown by the point ${\odot}$.

\begin{figure*}
{\includegraphics [width = 58mm] {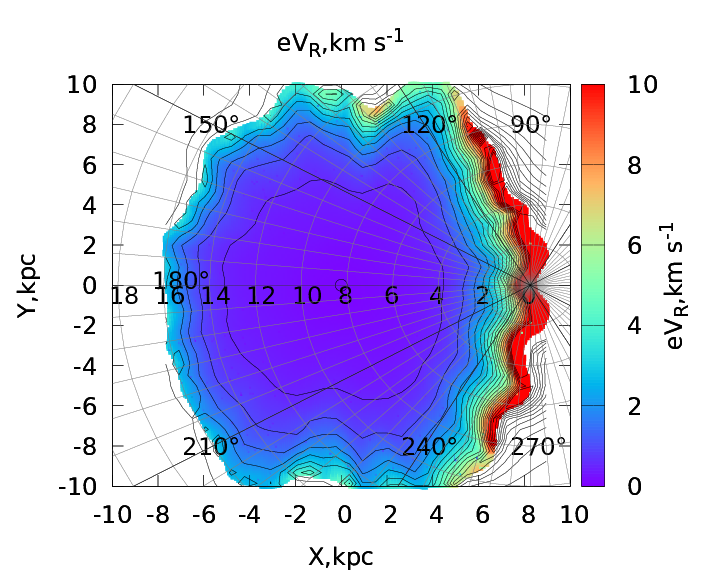}
 \includegraphics [width = 58mm] {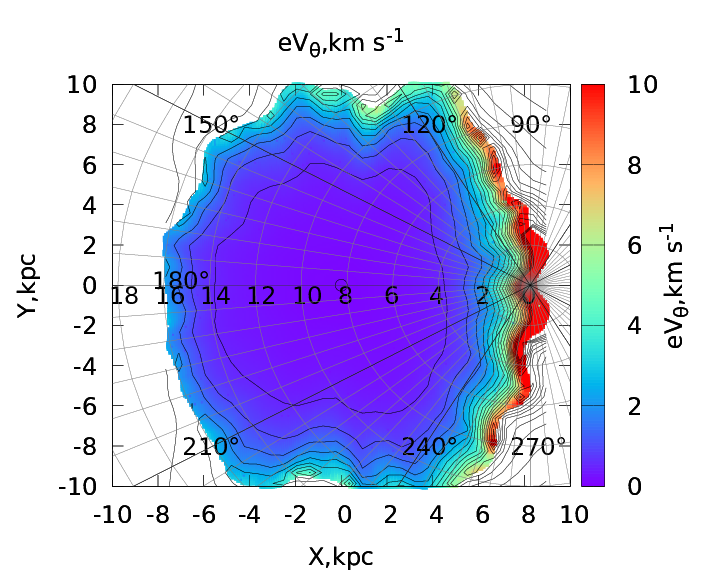}
 \includegraphics [width = 58mm] {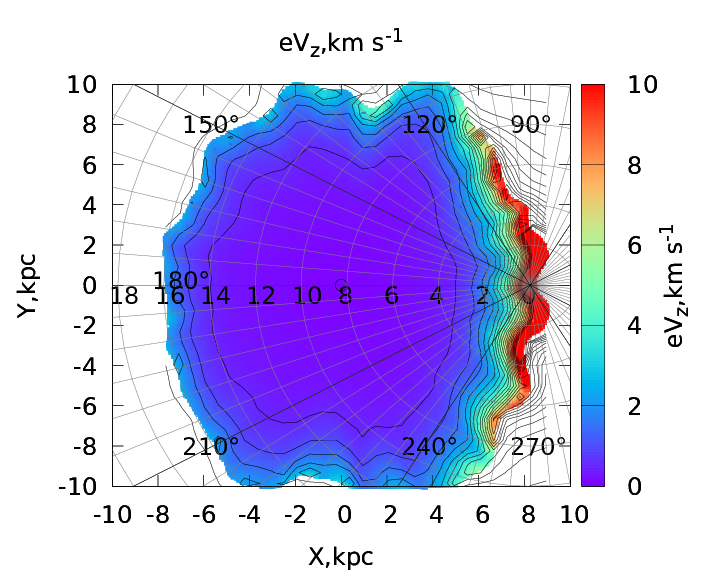}
 }
\caption{The random errors of the velocity components $V_R, V_\theta, V_Z$ as a function of the Galactic coordinates.}
\label{fig:eV}
\end{figure*}

\begin{figure*}
{\includegraphics [width = 58mm] {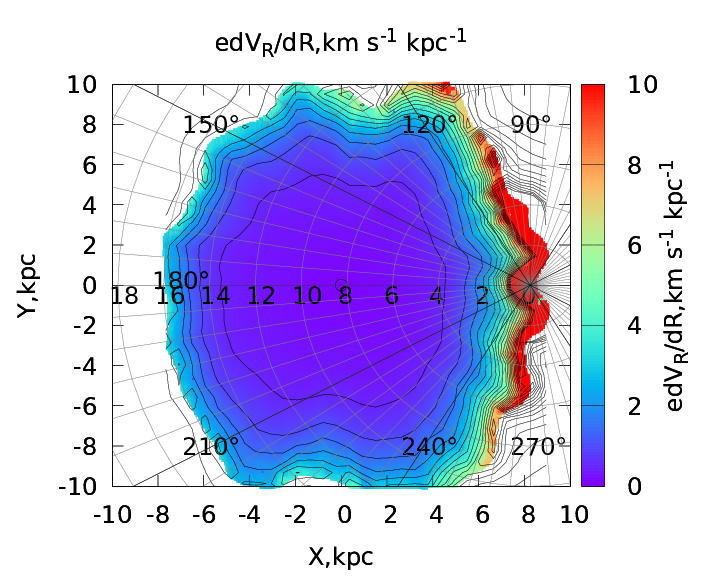}
 \includegraphics [width = 58mm] {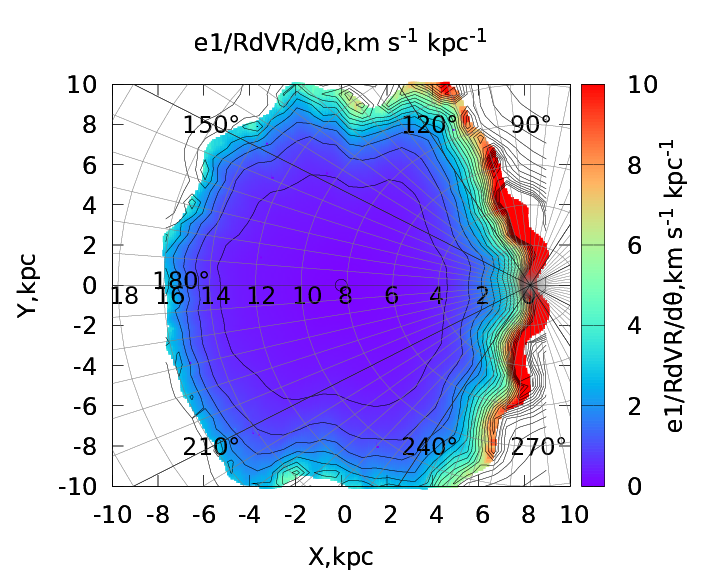}
 \includegraphics [width = 58mm] {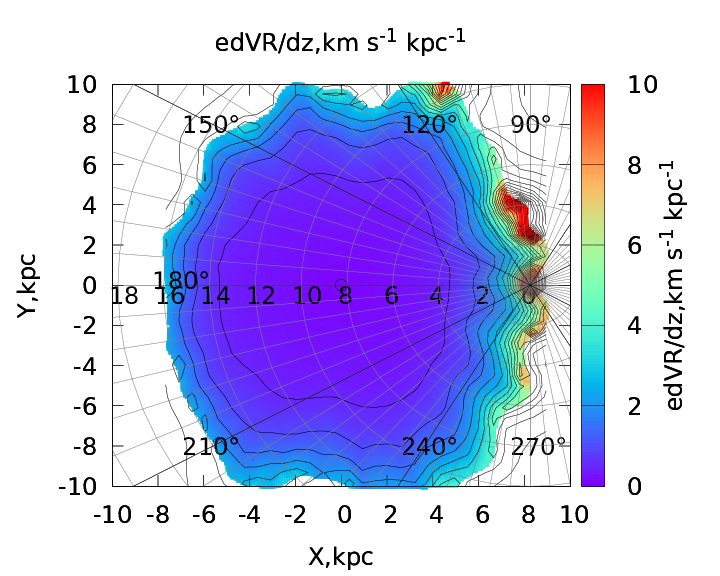}
 \includegraphics [width = 58mm] {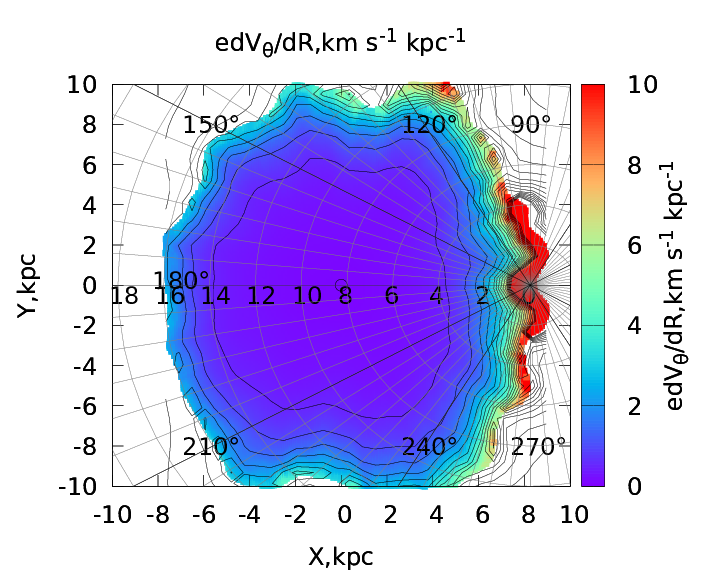}
 \includegraphics [width = 58mm] {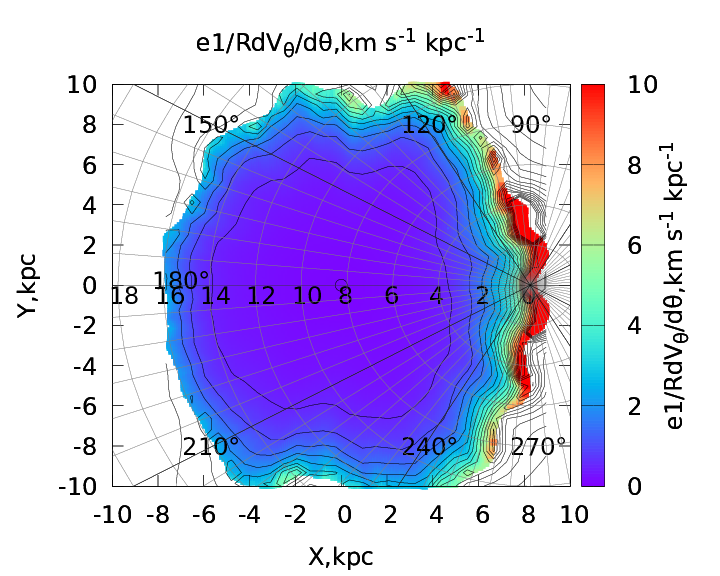}
 \includegraphics [width = 58mm] {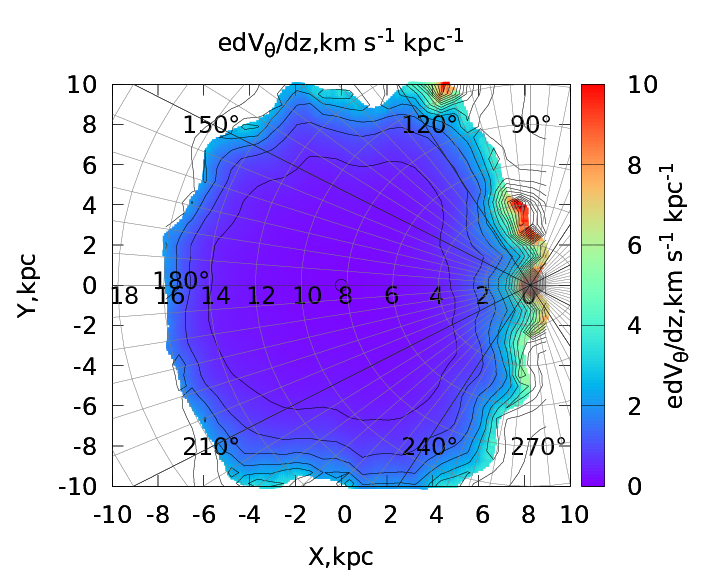}
 \includegraphics [width = 58mm] {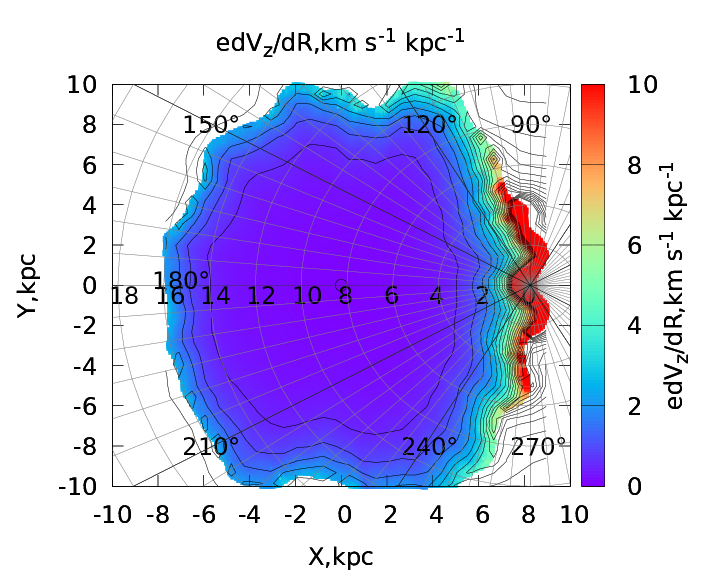}
 \includegraphics [width = 58mm] {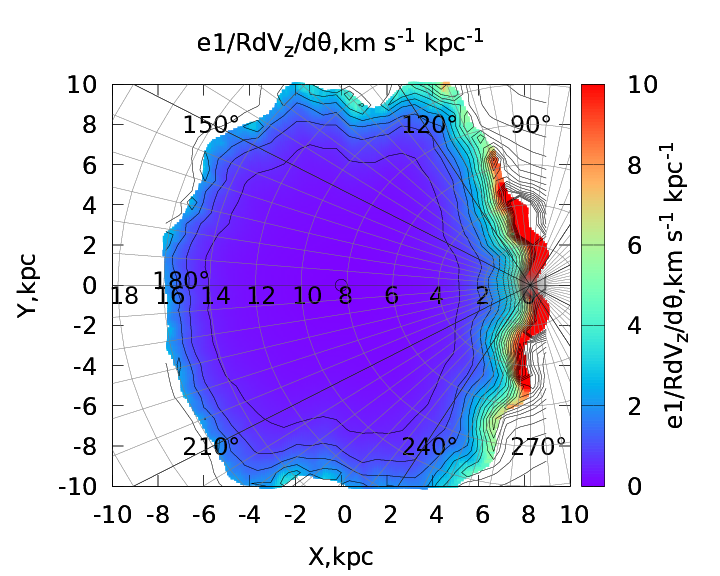}
 \includegraphics [width = 58mm] {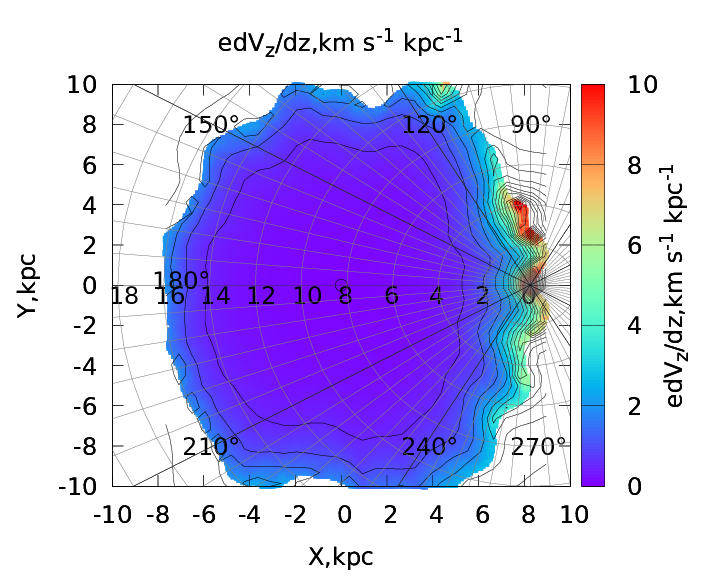}
 }
\caption{The random errors of the first order velocity derivatives estimates as a function of the Galactic coordinates.}
\label{fig:edV1}
\end{figure*}

\begin{figure*}
{\includegraphics [width = 58mm] {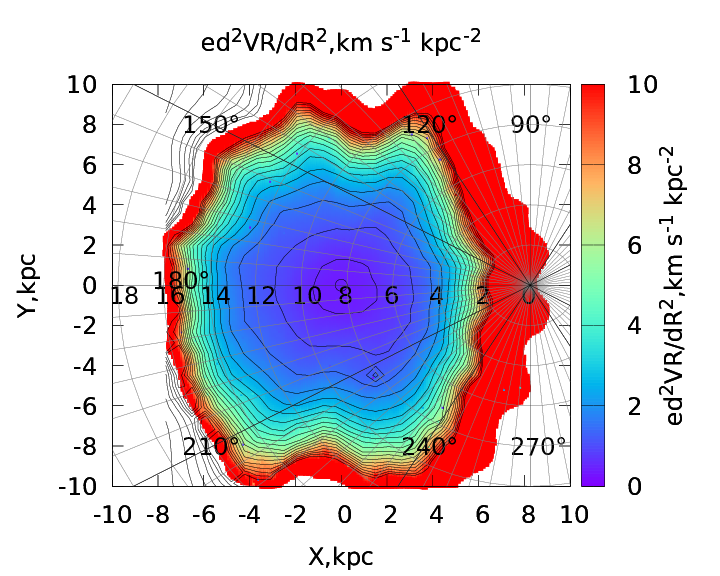}
 \includegraphics [width = 58mm] {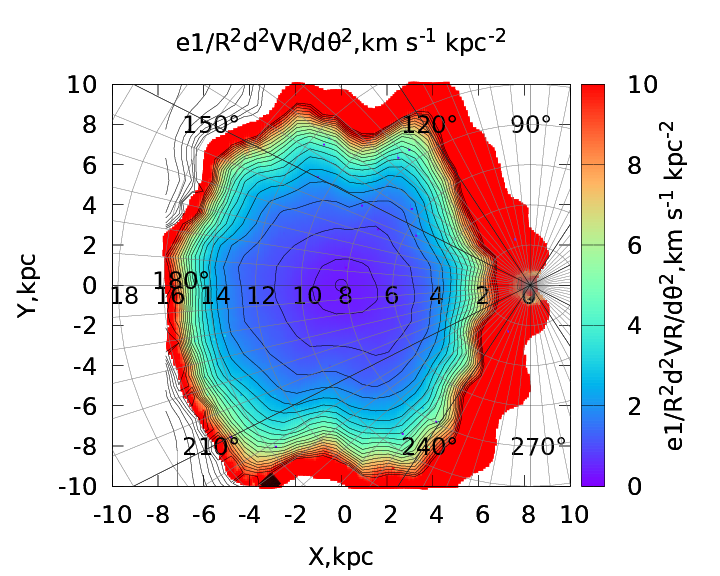}
 \includegraphics [width = 58mm] {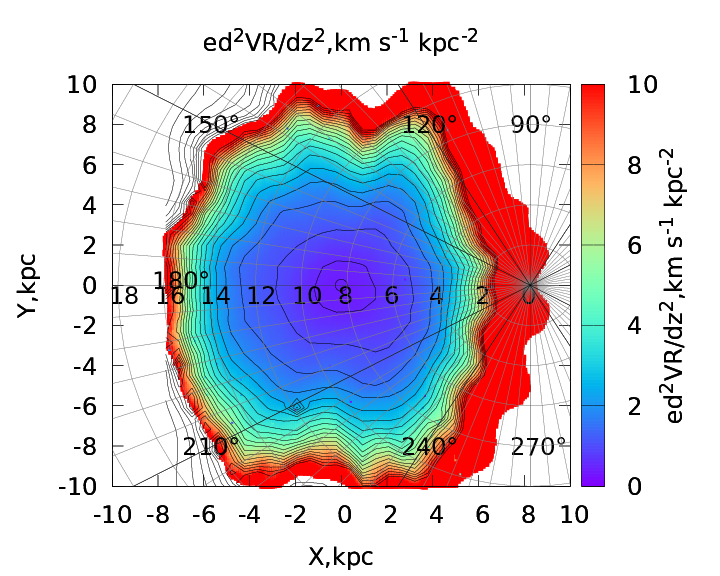}
 \includegraphics [width = 58mm] {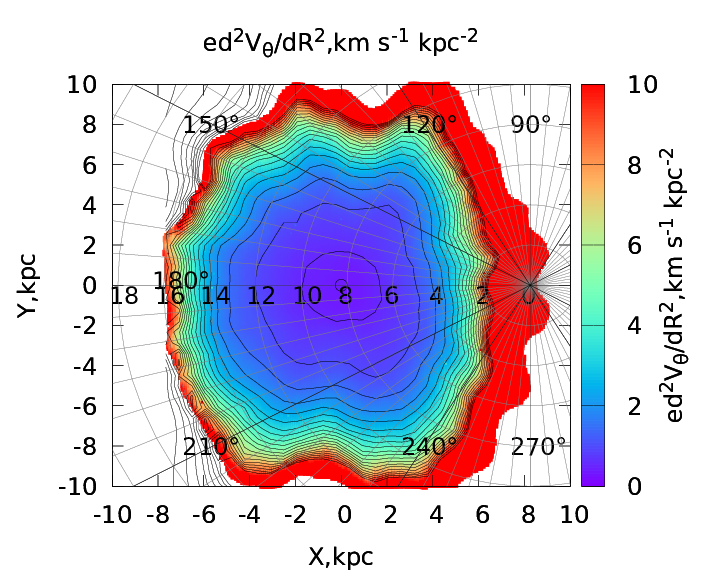}
 \includegraphics [width = 58mm] {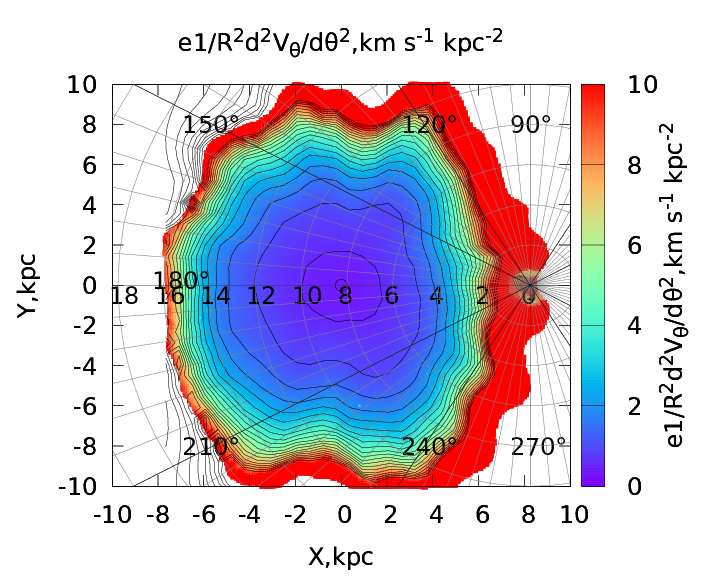}
 \includegraphics [width = 58mm] {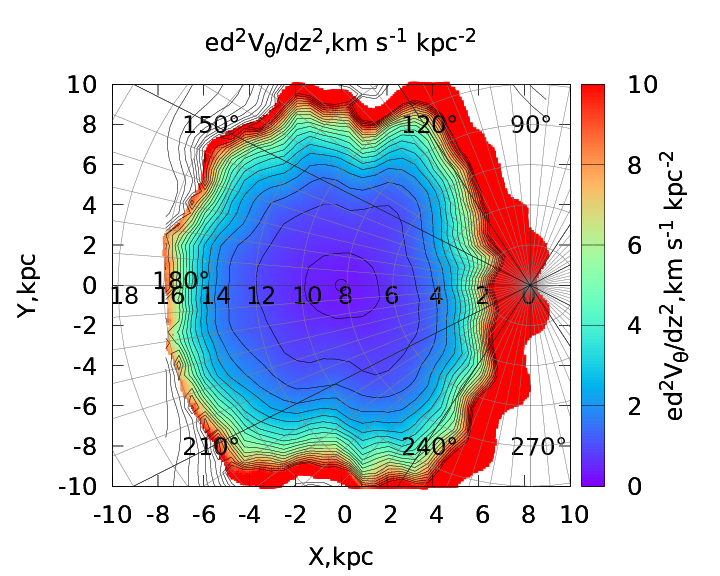}
 \includegraphics [width = 58mm] {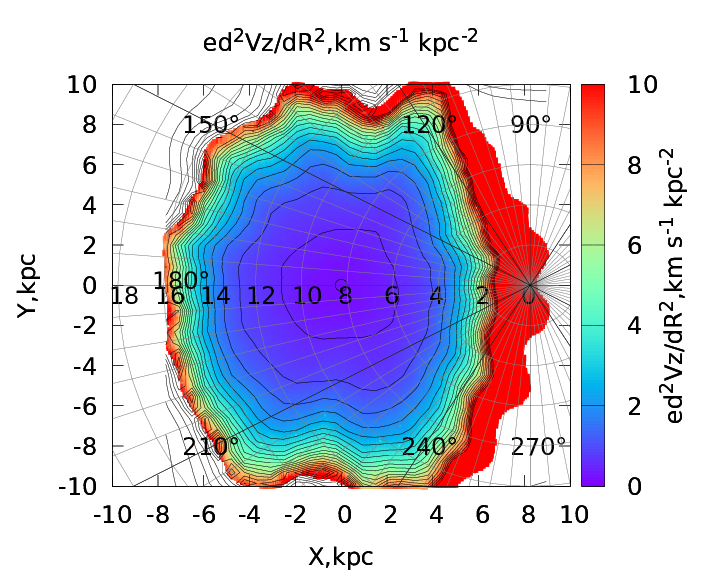}
 \includegraphics [width = 58mm] {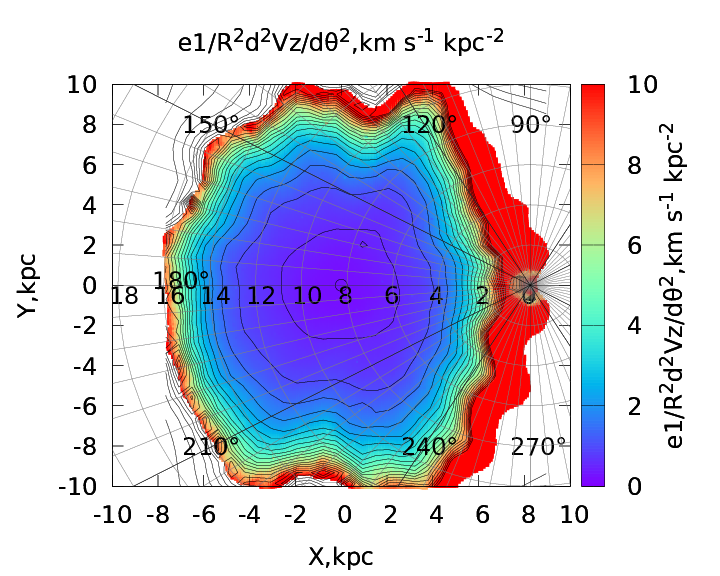}
 \includegraphics [width = 58mm] {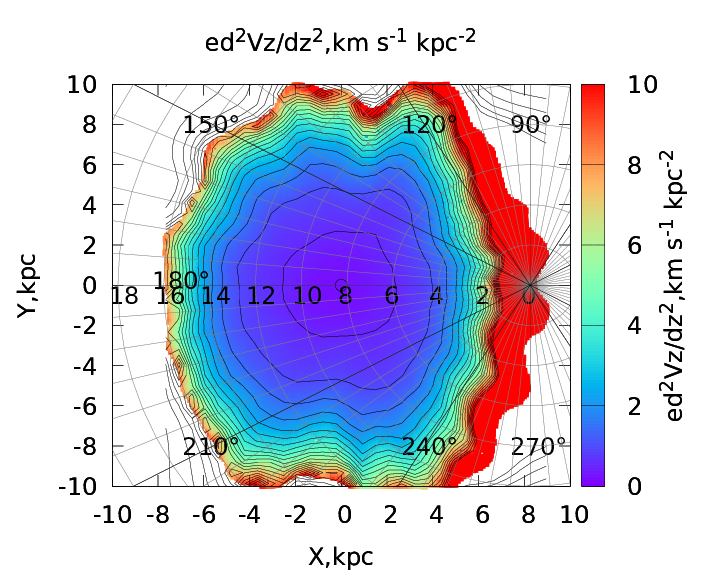}
 }
\caption{The random errors of the second order velocity derivatives estimates along directions $R, \theta, Z$ as a function of the Galactic coordinates.}
\label{fig:edV2}
\end{figure*}

\begin{figure*}
{\includegraphics [width = 58mm] {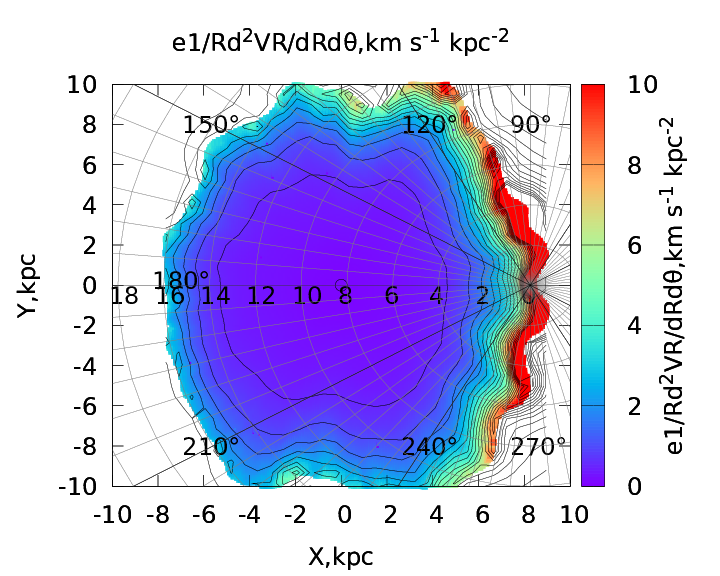}
 \includegraphics [width = 58mm] {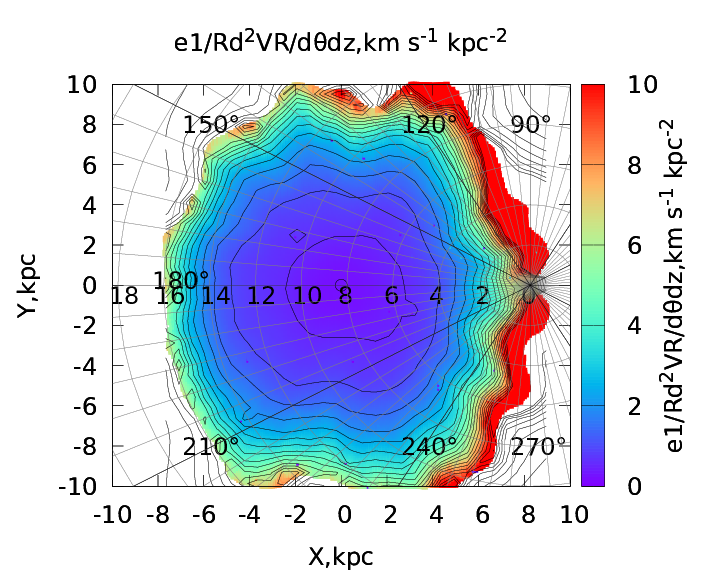}
 \includegraphics [width = 58mm] {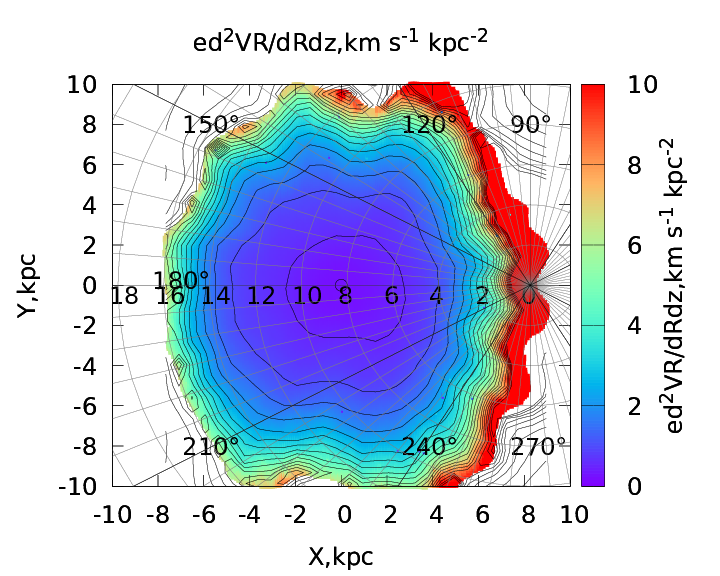}
 \includegraphics [width = 58mm] {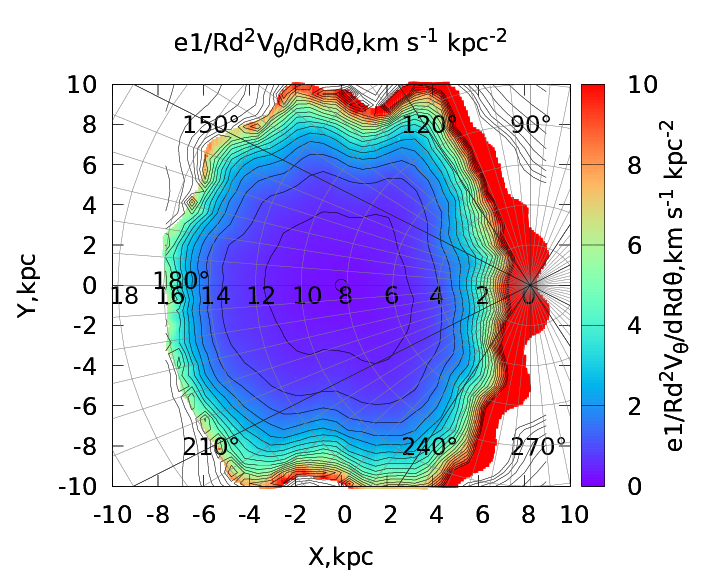}
 \includegraphics [width = 58mm] {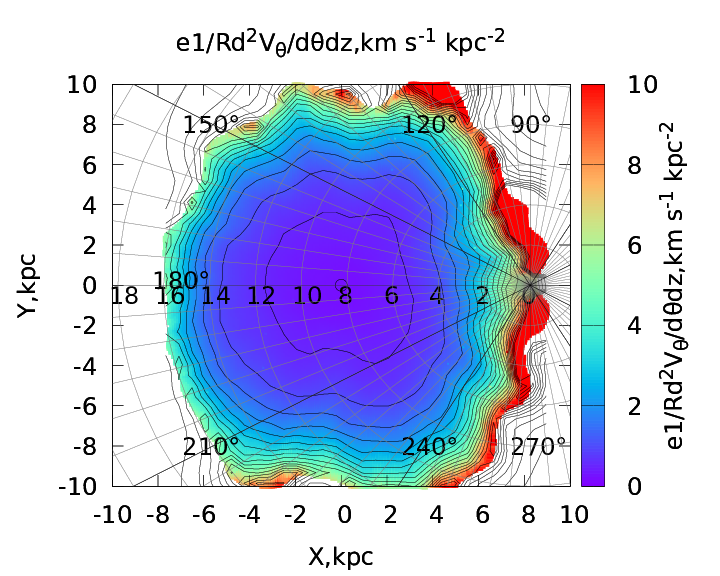}
 \includegraphics [width = 58mm] {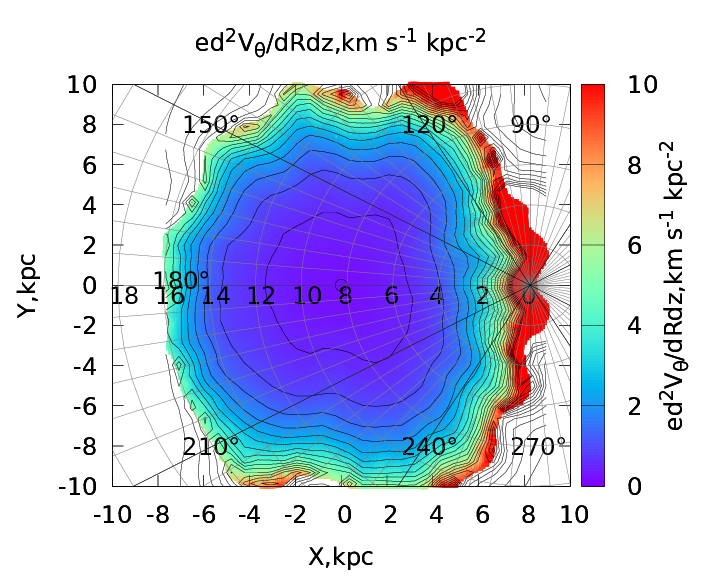}
 \includegraphics [width = 58mm] {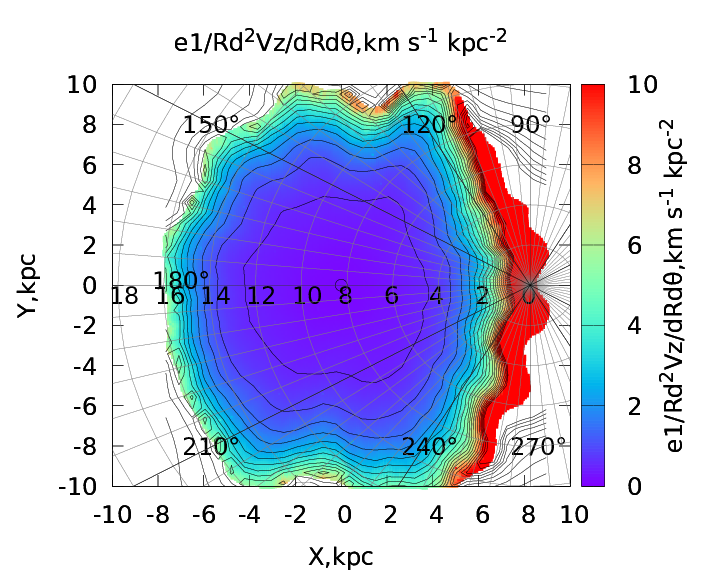}
 \includegraphics [width = 58mm] {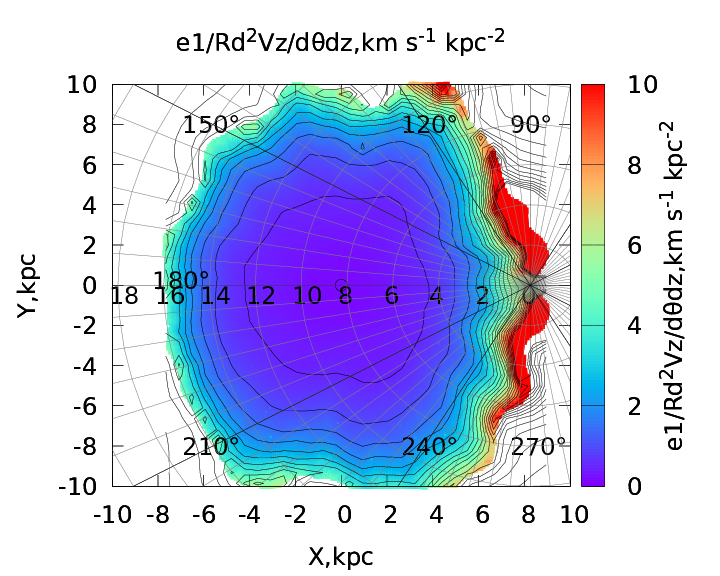}
 \includegraphics [width = 58mm] {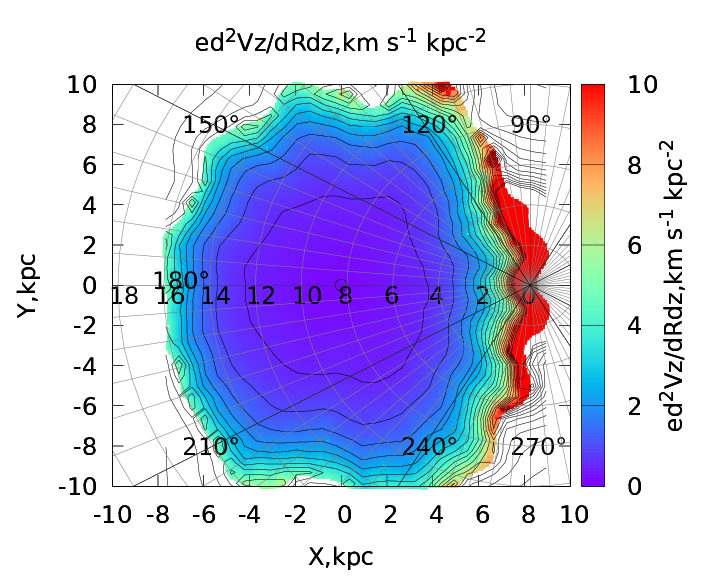}
 }
\caption{The random errors of the mixed second order velocity derivatives estimates as a function of the Galactic coordinates.}
\label{fig:edV2mix}
\end{figure*}

\begin{table*}
\caption{\label{tab:kinematic_param} The kinematic parameters for a stellar system close to the Sun with a radius of 1 kpc from equations (\ref{eq:derivativesVR}), (\ref{eq:derivativesVth}) and (\ref{eq:derivativesVz}) . The velocity components for the stellar system are in \kms; the first order velocity derivatives are in \kmskpc and the second order velocity derivatives are in \kmskpcc.}
\begin{center}
\begin{tabular}{|c|c|c|c|c|c|c|c|c|c|}
\hline
$V_R$ & $\frac{\partial V_R}{\partial R}$ & $\frac{1}{R}\frac{\partial V_R}{\partial \theta}$ & $\frac{\partial V_R}{\partial Z}$ & $\frac{\partial^2 V_R}{\partial R^2}$ & $\frac{1}{R}\frac{\partial^2 V_R}{\partial R \partial \theta}$ &  $\frac{\partial^2 V_R}{\partial R \partial Z}$ & $\frac{1}{R^2}\frac{\partial^2 V_R}{\partial \theta^2}$ & $\frac{1}{R}\frac{\partial^2 V_R}{\partial \theta \partial Z}$ & $\frac{\partial^2 V_R}{\partial Z^2}$ \\
\hline
 0.60$\pm$0.09 & -5.11$\pm$0.08 & 0.40$\pm$0.08 & 2.53$\pm$0.08 & 0.66$\pm$0.36 & -0.95$\pm$0.20 & 1.13$\pm$0.21 & 0.05$\pm$0.35  & -0.06$\pm$0.21 & 10.11$\pm$0.37 \\
\hline
$V_\theta$ & $\frac{1}{R}\frac{\partial V_\theta}{\partial R}$ & $\frac{1}{R}\frac{\partial V_\theta}{\partial \theta}$ & $\frac{\partial V_\theta}{\partial Z}$  & $\frac{\partial^2 V_\theta}{\partial R^2}$ & $\frac{1}{R}\frac{\partial^2 V_\theta}{\partial R \partial \theta}$ & $\frac{\partial^2 V_\theta}{\partial R \partial Z}$ & $\frac{1}{R^2}\frac{\partial^2 V_\theta}{\partial \theta^2}$ & $\frac{1}{R}\frac{\partial^2 V_\theta}{\partial \theta \partial Z}$ & $\frac{\partial^2 V_\theta}{\partial Z^2}$ \\
\hline
 -232.19$\pm$0.08 & 1.00$\pm$0.07 & -0.51$\pm$0.07 & 3.59$\pm$0.07 & 4.89 $\pm$0.32 & -0.36$\pm$0.18 & -4.216$\pm$0.19 & 0.47$\pm$0.32 & 0.25$\pm$0.19 & 53.87$\pm$0.33\\ 
 \hline
 $V_Z$ & $\frac{\partial V_Z}{\partial R}$ & $\frac{1}{R}\frac{\partial V_Z}{\partial \theta}$ & $\frac{\partial V_Z}{\partial Z}$ & $\frac{\partial^2 V_Z}{\partial R^2}$ & $\frac{1}{R}\frac{\partial^2 V_Z}{\partial R \partial \theta}$ &  $\frac{\partial^2 V_Z}{\partial R \partial Z}$ & $\frac{1}{R^2}\frac{\partial^2 V_Z}{\partial \theta^2}$ & $\frac{1}{R}\frac{\partial^2 V_Z}{\partial \theta \partial Z}$ & $\frac{\partial^2 V_Z}{\partial Z^2}$\\
\hline
 0.76$\pm$0.04 & 0.35$\pm$0.04 & -0.29$\pm$0.04 & -0.23$\pm$0.04 & 0.37$\pm$0.18 & -0.21$\pm$0.10 & -0.65$\pm$0.11 & 0.04$\pm$0.18 & 0.10$\pm$0.11 & -1.46$\pm$0.19\\
\hline
\end{tabular}
\end{center}
\end{table*}

\begin{table*}
\caption{\label{tab:Oort_constant} The Oort's  constants in \kmskpc for a stellar system close to the Sun with a radius of 1 kpc from equations (\ref{eq:Oort_A}),(\ref{eq:Oort_B}),(\ref{eq:Oort_C}) and (\ref{eq:Oort_K})} 
 \begin{center}
\begin{tabular}{|c|c|c|c|}
\hline
 \hline
 $A$ & $B$ & $C$ & $K$\\
\hline
 14.79$\pm$0.11 & -13.73$\pm$0.11 & -2.25$\pm$0.11 & -2.74$\pm$0.11\\
\hline
\end{tabular}
\end{center}
\end{table*}

\begin{table}
\caption{\label{tab:galaxy_curve_1} The values of the Galaxy rotation velocity are obtained by averaging the corresponding data in the range of angles $\theta$ from $150^\circ$ to $210^\circ$ ($\pm 30^{\circ}$ about the direction of the Galactic Centre -- the Sun -- the Galactic anticentre) in \kms as functions of the Galactocentric coordinates $R$ in kpc.}
\begin{center}
\begin{tabular}{|l|c|c|c|c|}
\hline
R & $\overline{-V_{\theta}}$(\ref{eq:gal_vel})& $-V_{\theta}$ (\ref{eq:derivativesVth})  & $V_{rotOM}$ (\ref{eq:Vrot}) & $V_{rotOM-\frac{dVR}{d\theta}}$ (\ref{eq:Vrot_dVR})\\
\hline
4.07 & 187.33$\pm$19.60 & 193.40$\pm$0.47 & 171.03$\pm$0.42 & 190.84$\pm$0.62\\
4.16 & 189.16$\pm$18.86 & 195.05$\pm$0.44 & 173.72$\pm$0.40 & 192.43$\pm$0.59\\
4.25 & 191.00$\pm$18.19 & 196.76$\pm$0.42 & 176.30$\pm$0.38 & 194.36$\pm$0.56\\
4.35 & 192.92$\pm$17.50 & 198.55$\pm$0.39 & 179.01$\pm$0.36 & 196.18$\pm$0.53\\
4.45 & 194.79$\pm$16.82 & 200.43$\pm$0.37 & 181.61$\pm$0.35 & 197.77$\pm$0.51\\
4.54 & 196.37$\pm$16.27 & 202.14$\pm$0.36 & 184.16$\pm$0.33 & 199.70$\pm$0.49\\
4.64 & 198.12$\pm$15.63 & 204.06$\pm$0.34 & 186.66$\pm$0.32 & 201.48$\pm$0.46\\
4.74 & 199.79$\pm$15.03 & 205.92$\pm$0.32 & 189.10$\pm$0.30 & 203.30$\pm$0.44\\
4.84 & 201.35$\pm$14.49 & 207.62$\pm$0.30 & 191.40$\pm$0.29 & 204.90$\pm$0.42\\
4.94 & 202.88$\pm$14.02 & 209.36$\pm$0.29 & 193.81$\pm$0.28 & 206.35$\pm$0.41\\
5.04 & 204.31$\pm$13.62 & 211.01$\pm$0.28 & 195.87$\pm$0.27 & 207.70$\pm$0.39\\
5.14 & 205.80$\pm$13.09 & 212.60$\pm$0.26 & 197.97$\pm$0.26 & 208.90$\pm$0.37\\
5.24 & 207.17$\pm$12.60 & 214.03$\pm$0.25 & 199.71$\pm$0.25 & 210.09$\pm$0.36\\
5.33 & 208.29$\pm$12.28 & 215.24$\pm$0.24 & 201.21$\pm$0.24 & 211.22$\pm$0.35\\
5.43 & 209.45$\pm$11.97 & 216.45$\pm$0.23 & 202.72$\pm$0.23 & 212.25$\pm$0.34\\
5.53 & 210.57$\pm$11.69 & 217.57$\pm$0.23 & 204.18$\pm$0.23 & 213.33$\pm$0.32\\
5.63 & 211.59$\pm$11.46 & 218.57$\pm$0.22 & 205.69$\pm$0.22 & 214.21$\pm$0.31\\
5.73 & 212.53$\pm$11.26 & 219.41$\pm$0.21 & 206.95$\pm$0.21 & 215.11$\pm$0.30\\
5.83 & 213.43$\pm$11.07 & 220.25$\pm$0.20 & 208.26$\pm$0.21 & 216.08$\pm$0.29\\
5.93 & 214.32$\pm$10.86 & 221.07$\pm$0.19 & 209.67$\pm$0.20 & 216.96$\pm$0.28\\
6.03 & 215.10$\pm$10.68 & 221.83$\pm$0.18 & 210.75$\pm$0.20 & 217.73$\pm$0.28\\
6.13 & 215.82$\pm$10.55 & 222.52$\pm$0.18 & 212.24$\pm$0.19 & 218.39$\pm$0.27\\
6.22 & 216.47$\pm$10.38 & 223.16$\pm$0.17 & 213.47$\pm$0.19 & 219.15$\pm$0.26\\
6.33 & 217.17$\pm$10.25 & 223.79$\pm$0.17 & 214.73$\pm$0.18 & 219.87$\pm$0.25\\
6.43 & 217.84$\pm$10.05 & 224.36$\pm$0.16 & 215.80$\pm$0.18 & 220.43$\pm$0.25\\
6.52 & 218.43$\pm$9.88 & 224.89$\pm$0.16 & 216.87$\pm$0.17 & 220.94$\pm$0.24\\
6.62 & 219.08$\pm$9.70 & 225.45$\pm$0.15 & 217.86$\pm$0.17 & 221.66$\pm$0.24\\
6.72 & 219.68$\pm$9.47 & 225.97$\pm$0.15 & 218.78$\pm$0.16 & 222.15$\pm$0.23\\
6.82 & 220.27$\pm$9.23 & 226.50$\pm$0.15 & 219.53$\pm$0.16 & 222.86$\pm$0.23\\
6.92 & 220.82$\pm$9.02 & 226.99$\pm$0.14 & 220.29$\pm$0.16 & 223.43$\pm$0.22\\
7.02 & 221.34$\pm$8.87 & 227.49$\pm$0.14 & 221.05$\pm$0.15 & 223.97$\pm$0.22\\
7.11 & 221.79$\pm$8.73 & 227.90$\pm$0.14 & 221.61$\pm$0.15 & 224.44$\pm$0.21\\
7.22 & 222.31$\pm$8.57 & 228.40$\pm$0.14 & 222.37$\pm$0.15 & 224.90$\pm$0.21\\
7.32 & 222.75$\pm$8.43 & 228.85$\pm$0.14 & 223.04$\pm$0.15 & 225.41$\pm$0.21\\
7.42 & 223.20$\pm$8.31 & 229.31$\pm$0.13 & 223.63$\pm$0.14 & 225.91$\pm$0.21\\
7.52 & 223.57$\pm$8.19 & 229.65$\pm$0.13 & 224.29$\pm$0.14 & 226.23$\pm$0.21\\
7.61 & 223.94$\pm$8.10 & 229.97$\pm$0.13 & 225.04$\pm$0.14 & 226.60$\pm$0.21\\
7.71 & 224.25$\pm$8.00 & 230.22$\pm$0.14 & 225.73$\pm$0.14 & 226.86$\pm$0.21\\
7.81 & 224.56$\pm$7.93 & 230.45$\pm$0.14 & 226.44$\pm$0.14 & 227.06$\pm$0.21\\
7.91 & 224.82$\pm$7.87 & 230.56$\pm$0.14 & 227.14$\pm$0.14 & 227.26$\pm$0.21\\
8.01 & 225.07$\pm$7.85 & 230.66$\pm$0.14 & 227.89$\pm$0.14 & 227.47$\pm$0.21\\
8.11 & 225.29$\pm$7.81 & 230.71$\pm$0.14 & 228.52$\pm$0.14 & 227.57$\pm$0.22\\
8.21 & 225.52$\pm$7.79 & 230.70$\pm$0.14 & 229.37$\pm$0.14 & 227.76$\pm$0.22\\
8.30 & 225.67$\pm$7.75 & 230.66$\pm$0.15 & 229.95$\pm$0.14 & 227.79$\pm$0.22\\
8.40 & 225.85$\pm$7.72 & 230.60$\pm$0.15 & 230.51$\pm$0.14 & 227.98$\pm$0.22\\
8.50 & 226.01$\pm$7.66 & 230.54$\pm$0.15 & 231.12$\pm$0.15 & 228.01$\pm$0.22\\
8.60 & 226.16$\pm$7.60 & 230.45$\pm$0.15 & 231.37$\pm$0.15 & 228.20$\pm$0.23\\
8.70 & 226.28$\pm$7.52 & 230.38$\pm$0.16 & 232.06$\pm$0.15 & 228.32$\pm$0.23\\
8.81 & 226.37$\pm$7.46 & 230.31$\pm$0.16 & 232.43$\pm$0.15 & 228.23$\pm$0.23\\
8.90 & 226.47$\pm$7.36 & 230.23$\pm$0.16 & 232.59$\pm$0.15 & 228.24$\pm$0.24\\
9.00 & 226.53$\pm$7.32 & 230.14$\pm$0.16 & 232.81$\pm$0.15 & 228.18$\pm$0.24\\
9.10 & 226.59$\pm$7.22 & 230.07$\pm$0.17 & 233.12$\pm$0.16 & 228.22$\pm$0.25\\
9.20 & 226.63$\pm$7.20 & 229.98$\pm$0.17 & 233.26$\pm$0.16 & 228.27$\pm$0.25\\
9.30 & 226.65$\pm$7.12 & 229.88$\pm$0.17 & 233.56$\pm$0.16 & 228.25$\pm$0.26\\
9.40 & 226.68$\pm$7.06 & 229.74$\pm$0.17 & 233.46$\pm$0.16 & 228.19$\pm$0.26\\
9.50 & 226.66$\pm$6.96 & 229.58$\pm$0.18 & 233.67$\pm$0.16 & 228.11$\pm$0.26\\
9.60 & 226.65$\pm$6.93 & 229.43$\pm$0.18 & 233.79$\pm$0.17 & 228.24$\pm$0.27\\
9.70 & 226.61$\pm$6.83 & 229.24$\pm$0.18 & 233.75$\pm$0.17 & 228.16$\pm$0.27\\
9.80 & 226.54$\pm$6.89 & 229.08$\pm$0.18 & 233.77$\pm$0.17 & 228.13$\pm$0.28\\
9.90 & 226.43$\pm$6.91 & 228.90$\pm$0.19 & 233.73$\pm$0.17 & 227.98$\pm$0.29\\
\hline
\end{tabular}
\end{center}
\end{table}

\begin{table}
\caption{\label{tab:galaxy_curve_2} The values of the Galaxy rotation velocity are obtained by averaging the corresponding data in the range of angles $\theta$ from $150^\circ$ to $210^\circ$ ($\pm 30^{\circ}$ about the direction of the Galactic Centre -- the Sun -- the Galactic anticentre) in \kms as functions of the Galactocentric coordinates $R$ in kpc.}
\begin{center}
\begin{tabular}{|l|c|c|c|c|c|c|}
\hline
R & $\overline{-V_{\theta}}$(\ref{eq:gal_vel})& $-V_{\theta}$ (\ref{eq:derivativesVth})  & $V_{rotOM}$ (\ref{eq:Vrot}) & $V_{rotOM-\frac{dVR}{d\theta}}$ (\ref{eq:Vrot_dVR})\\
\hline
10.00 & 226.28$\pm$6.99 & 228.72$\pm$0.19 & 233.49$\pm$0.18 & 227.87$\pm$0.29\\
10.10 & 226.10$\pm$7.05 & 228.52$\pm$0.19 & 233.47$\pm$0.18 & 227.79$\pm$0.30\\
10.20 & 225.91$\pm$7.13 & 228.28$\pm$0.20 & 233.21$\pm$0.19 & 227.69$\pm$0.30\\
10.30 & 225.68$\pm$7.21 & 227.99$\pm$0.20 & 232.66$\pm$0.19 & 227.34$\pm$0.31\\
10.39 & 225.46$\pm$7.33 & 227.68$\pm$0.20 & 232.47$\pm$0.19 & 227.36$\pm$0.32\\
10.49 & 225.24$\pm$7.46 & 227.32$\pm$0.21 & 232.06$\pm$0.20 & 227.16$\pm$0.32\\
10.60 & 224.99$\pm$7.63 & 226.90$\pm$0.21 & 231.30$\pm$0.20 & 227.01$\pm$0.33\\
10.69 & 224.72$\pm$7.78 & 226.52$\pm$0.22 & 230.45$\pm$0.21 & 226.47$\pm$0.34\\
10.80 & 224.46$\pm$7.95 & 226.10$\pm$0.22 & 229.89$\pm$0.21 & 226.28$\pm$0.35\\
10.89 & 224.23$\pm$8.13 & 225.80$\pm$0.23 & 229.29$\pm$0.22 & 226.14$\pm$0.35\\
10.99 & 223.99$\pm$8.32 & 225.51$\pm$0.24 & 228.81$\pm$0.22 & 226.08$\pm$0.36\\
11.10 & 223.77$\pm$8.54 & 225.24$\pm$0.25 & 227.99$\pm$0.23 & 225.81$\pm$0.37\\
11.19 & 223.62$\pm$8.77 & 225.05$\pm$0.25 & 227.25$\pm$0.24 & 225.50$\pm$0.38\\
11.29 & 223.48$\pm$9.01 & 224.88$\pm$0.26 & 226.60$\pm$0.24 & 225.35$\pm$0.39\\
11.39 & 223.38$\pm$9.28 & 224.77$\pm$0.27 & 226.05$\pm$0.25 & 225.27$\pm$0.40\\
11.49 & 223.33$\pm$9.55 & 224.71$\pm$0.28 & 225.19$\pm$0.26 & 225.12$\pm$0.41\\
11.59 & 223.31$\pm$9.88 & 224.71$\pm$0.29 & 224.61$\pm$0.27 & 225.19$\pm$0.43\\
11.69 & 223.31$\pm$10.18 & 224.78$\pm$0.30 & 223.66$\pm$0.28 & 225.14$\pm$0.44\\
11.79 & 223.35$\pm$10.44 & 224.88$\pm$0.32 & 223.42$\pm$0.28 & 225.50$\pm$0.45\\
11.88 & 223.46$\pm$10.73 & 225.07$\pm$0.32 & 222.40$\pm$0.29 & 225.33$\pm$0.46\\
11.98 & 223.55$\pm$11.07 & 225.19$\pm$0.34 & 221.63$\pm$0.30 & 225.46$\pm$0.48\\
12.08 & 223.58$\pm$11.45 & 225.34$\pm$0.34 & 221.32$\pm$0.31 & 226.03$\pm$0.49\\
12.18 & 223.67$\pm$11.84 & 225.47$\pm$0.36 & 220.93$\pm$0.32 & 226.24$\pm$0.51\\
12.29 & 223.76$\pm$12.26 & 225.65$\pm$0.37 & 219.88$\pm$0.33 & 226.24$\pm$0.53\\
12.38 & 223.81$\pm$12.63 & 225.68$\pm$0.38 & 219.69$\pm$0.34 & 226.30$\pm$0.54\\
12.49 & 223.90$\pm$13.09 & 225.73$\pm$0.40 & 218.87$\pm$0.36 & 226.27$\pm$0.56\\
12.58 & 223.90$\pm$13.56 & 225.81$\pm$0.40 & 218.04$\pm$0.37 & 226.07$\pm$0.58\\
12.67 & 223.89$\pm$13.84 & 225.76$\pm$0.42 & 217.32$\pm$0.38 & 225.96$\pm$0.60\\
12.77 & 223.95$\pm$14.32 & 225.79$\pm$0.43 & 217.13$\pm$0.39 & 226.54$\pm$0.62\\
12.86 & 223.89$\pm$14.67 & 225.64$\pm$0.44 & 216.63$\pm$0.41 & 226.50$\pm$0.64\\
12.96 & 223.86$\pm$15.18 & 225.56$\pm$0.46 & 215.89$\pm$0.42 & 225.97$\pm$0.67\\
13.06 & 223.73$\pm$15.70 & 225.42$\pm$0.48 & 215.43$\pm$0.44 & 225.50$\pm$0.69\\
13.16 & 223.65$\pm$16.24 & 225.29$\pm$0.50 & 214.30$\pm$0.46 & 225.34$\pm$0.72\\
13.26 & 223.49$\pm$16.78 & 225.18$\pm$0.52 & 214.08$\pm$0.48 & 224.89$\pm$0.75\\
13.35 & 223.36$\pm$17.25 & 225.11$\pm$0.54 & 213.54$\pm$0.49 & 225.26$\pm$0.78\\
13.44 & 223.19$\pm$17.74 & 225.11$\pm$0.56 & 213.08$\pm$0.52 & 225.13$\pm$0.81\\
13.54 & 223.02$\pm$18.26 & 225.05$\pm$0.58 & 212.73$\pm$0.54 & 225.30$\pm$0.85\\
13.62 & 222.87$\pm$18.74 & 224.93$\pm$0.61 & 212.31$\pm$0.56 & 224.91$\pm$0.88\\
13.71 & 222.68$\pm$19.22 & 224.87$\pm$0.64 & 210.72$\pm$0.59 & 225.03$\pm$0.93\\
13.81 & 222.46$\pm$19.79 & 224.62$\pm$0.67 & 209.67$\pm$0.62 & 224.53$\pm$0.97\\
13.90 & 222.27$\pm$20.26 & 224.35$\pm$0.71 & 208.96$\pm$0.65 & 224.16$\pm$1.03\\
13.99 & 222.06$\pm$20.72 & 224.14$\pm$0.75 & 208.66$\pm$0.69 & 222.94$\pm$1.08\\
14.08 & 221.83$\pm$21.30 & 223.85$\pm$0.79 & 206.84$\pm$0.73 & 223.54$\pm$1.15\\
14.17 & 221.54$\pm$21.89 & 223.62$\pm$0.83 & 205.49$\pm$0.77 & 223.53$\pm$1.22\\
14.26 & 221.27$\pm$22.42 & 223.35$\pm$0.89 & 205.42$\pm$0.81 & 223.30$\pm$1.29\\
14.35 & 220.97$\pm$22.90 & 223.01$\pm$0.95 & 203.12$\pm$0.86 & 222.51$\pm$1.38\\
14.44 & 220.68$\pm$23.49 & 222.61$\pm$1.00 & 201.34$\pm$0.92 & 222.12$\pm$1.47\\
14.53 & 220.35$\pm$24.06 & 222.34$\pm$1.07 & 201.07$\pm$0.97 & 221.19$\pm$1.56\\
14.61 & 220.08$\pm$24.49 & 222.05$\pm$1.14 & 201.78$\pm$1.03 & 222.79$\pm$1.66\\
14.70 & 219.72$\pm$25.02 & 221.79$\pm$1.21 & 202.14$\pm$1.09 & 219.93$\pm$1.77\\
14.79 & 219.34$\pm$25.43 & 221.54$\pm$1.30 & 199.07$\pm$1.17 & 218.78$\pm$1.90\\
14.89 & 219.00$\pm$25.91 & 221.33$\pm$1.39 & 199.35$\pm$1.25 & 219.62$\pm$2.04\\
14.98 & 218.72$\pm$26.14 & 220.82$\pm$1.48 & 198.32$\pm$1.35 & 218.25$\pm$2.20\\
15.08 & 218.47$\pm$26.47 & 220.31$\pm$1.59 & 198.28$\pm$1.45 & 220.10$\pm$2.37\\
15.17 & 218.04$\pm$27.09 & 219.99$\pm$1.70 & 195.00$\pm$1.56 & 217.52$\pm$2.56\\
15.26 & 217.65$\pm$27.70 & 219.61$\pm$1.81 & 195.53$\pm$1.68 & 218.09$\pm$2.77\\
15.36 & 217.29$\pm$28.21 & 219.33$\pm$1.95 & 193.54$\pm$1.80 & 213.26$\pm$2.99\\
15.46 & 216.97$\pm$28.62 & 218.95$\pm$2.12 & 190.74$\pm$1.95 & 215.02$\pm$3.25\\
15.55 & 216.73$\pm$28.90 & 218.80$\pm$2.29 & 193.54$\pm$2.08 & 217.57$\pm$3.51\\
15.65 & 216.41$\pm$29.50 & 218.59$\pm$2.45 & 194.75$\pm$2.24 & 219.36$\pm$3.79\\
\hline
\end{tabular}
\end{center}
\end{table}

\end{document}